\documentclass[fleqn,usenatbib]{mnras}

\usepackage{newtxtext,newtxmath}

\usepackage[T1]{fontenc}
\usepackage{amsmath}
\usepackage{graphicx}
\usepackage{lastpage}
\usepackage[dvipsnames]{xcolor}   
\usepackage[super]{nth}
\usepackage{microtype}
\usepackage{booktabs, tabularx}
\usepackage{supertabular}
\usepackage{physics}
\usepackage[math]{cellspace}

\usepackage{hyperref}
\definecolor{royalblue}{rgb}{0.25, 0.41, 0.88}
\definecolor{darkviolet}{rgb}{0.58, 0.0, 0.83}

\newcommand{\ie}{{\it i.e. }}

\newcommand{\ddf}[3][]{\frac{\dd^{#1} #2}{\dd {#3}^{#1}}}

\title{The resilience of the Etherington--Hubble relation}

\author[F. Renzi, N. B. Hogg and W. Giar\`e]{
Fabrizio Renzi,$^{1}$\thanks{E-mail: renzi@lorentz.leidenuniv.nl }
Natalie B. Hogg,$^{2}$\thanks{E-mail: natalie.hogg@uam.es}
William Giar\`e,$^{3,4}$\thanks{E-mail: william.giare@uniroma1.it }
\\
$^{1}$ Lorentz Institute for Theoretical Physics, Leiden University, PO Box 9506, Leiden 2300 RA, The Netherlands\\
$^{2}$ Instituto de F\'isica Te\'orica UAM-CSIC, U. Aut\'onoma de Madrid, C/ Nicol\'as Cabrera 13-15, Cantoblanco, 28049, Madrid, Spain\\ 
$^{3}$ Galileo Galileo Institute for theoretical physics, Centro Nazionale INFN di studi avanzati, Largo Enrico Fermi 2,  I-50125, Firenze, Italy\\
$^{4}$ INFN Sezione di Roma, P.le A. Moro 2, I-00185, Roma, Italy}

\pubyear{2021}

\begin{document}
\label{firstpage}
\pagerange{\pageref{firstpage}--\pageref{lastpage}}
\maketitle

\begin{abstract} 
The Etherington reciprocity theorem, or distance duality relation (DDR), relates the mutual scaling of cosmic distances in any metric theory of gravity where photons are massless and propagate on null geodesics. In this paper, we make use of the DDR to build a consistency check based on its degeneracy with the Hubble constant, $H_0$. We parameterise the DDR using the form $\eta(z) = 1+ \epsilon z$, thus only allowing small deviations from its standard value. We use a combination of late time observational data to provide the first joint constraints on the Hubble parameter and $\epsilon$ with percentage accuracy: $H_0 = 68.6 \pm 2.5$ kms$^{-1}$Mpc$^{-1}$ and $\epsilon = 0.001^{+0.023}_{-0.026}$.  We build our consistency check using these constraints and compare them with the results obtained in extended cosmological models using cosmic microwave background data. We find that extensions to $\Lambda$CDM involving massive neutrinos and/or additional dark radiation are in perfect agreement with the DDR, while models with non-zero spatial curvature show a preference for DDR violation, \ie $\epsilon \ne 0 $ at the level of $\sim 1.5 \sigma$. Most importantly, we find a mild 2$\sigma$ discrepancy between the validity of the DDR and the latest publicly available Cepheid-calibrated SNIa constraint on $H_0$. We discuss the potential consequences of this for both the Etherington reciprocity theorem and the $H_0$ tension.
\end{abstract}


\section{Introduction}
The era of precision cosmology began with the launch of the COBE satellite in 1989, which measured both the near-perfect black body spectrum and the tiny temperature anisotropies of the cosmic microwave background (CMB) \citep{Fixsen:1993rd, Bennett:1996ce}. Since then, a vast array of surveys and experiments have probed nearly all epochs and scales of the Universe, converging on the standard $\Lambda$CDM model of cosmology. This model describes a Universe which is spatially flat (the curvature density $\Omega_k =0$), is dominated at late times by a cosmological constant $\Lambda$ and in which the majority of matter only interacts gravitationally. We call this type of matter cold dark matter (CDM). Furthermore, we assume that general relativity is the correct description of gravity, and that on the largest scales the Universe is homogeneous and isotropic. This \textit{cosmological principle} is  encapsulated in the construction of the Friedmann--Lema\^{i}tre--Robertson--Walker (FLRW) metric \citep{Weinberg}.

The Hubble expansion rate $H(z)$ is a fundamental quantity in any cosmological model based on a theory of gravity which uses the FLRW metric. There are a number of different ways that the value of the Hubble parameter at redshift zero, \ie the Hubble constant $H_0$, can be measured, some of which depend on the cosmological model chosen. With the increasing precision of the observational data at our disposal, it has become clear that a significant statistical tension exists between some of the different measurements of $H_0$ \citep{DiValentino:2015ola,Verde:2019ivm,Pogosian:2021mcs,Freedman:2021ahq,DiValentino:2020hov}. Specifically, when measurements are made by constructing a distance ladder to Type Ia supernovae (SNIa) using parallax distances and the period--luminosity relation of Cepheid variable stars, the Hubble constant is found to be $H_0 = 73.2 \pm 1.3$ kms$^{-1}$ Mpc$^{-1}$ (by the S$H_0$ES collaboration \citep{Riess:2020fzl}, R20 hereafter), which is in tension with the Planck value stemming from measurements of the temperature anisotropies and polarisation of the cosmic microwave background, $H_0 = 67.4 \pm 0.5$ kms$^{-1}$ Mpc$^{-1}$ \citep{Aghanim:2018eyx} (P18 hereafter), a cosmological-model-dependent result.

It is worth noting that while the S$H_0$ES collaboration uses light curves of Cepheid variable stars to anchor the SNIa and hence infer $H_0$, alternative methods of calibration, using either Mira variable stars or stars at the tip of the red giant branch (TRBG) have measured values of $H_0 = 73.6 \pm 3.9$ kms$^{-1}$Mpc$^{-1}$ \citep{Huang:2018dbn, Huang:2019yhh} and $H_0 = 69.6 \pm 1.9$ kms$^{-1}$Mpc$^{-1}$ respectively \citep{Freedman:2019jwv,Freedman:2021ahq}. In particular, the TRGB result shows a $\sim2\sigma$ tension with R20 that can be linked to an inconsistency in their measurements of distances to common SNIa hosts \citep{Efstathiou:2021ocp}. Some unaccounted-for systematics in the SNIa calibration could possibly resolve the moderate tension between these methods but a compelling answer is yet to be found \citep{Mortsell:2021nzg,Martinelli:2019krf}. 

Similar arguments can be applied to CMB observations, as either a modification of the recombination physics (by e.g. early dark energy \citep{Poulin:2018cxd, Niedermann:2020dwg, Freese:2021rjq} or particles beyond the Standard Model~\citep{DEramo:2018vss,Arias-Aragon:2020qtn,Arias-Aragon:2020shv,Giare:2020vzo})  or of the expansion rate (by e.g. interacting dark energy \citep{Salvatelli2014,  Martinelli:2019dau}, dynamical dark energy \citep{Zhao:2017cud, Bonilla:2020wbn} or modified gravity \citep{Raveri:2019mxg, Peirone:2019aua}) can lead to significant variations in the value of $H_0$ inferred by Planck. There have been many efforts in this direction to find a solution to the Hubble tension, but so far these have been largely unsuccessful (see e.g. \cite{Knox:2019rjx,DiValentino:2021izs,Abdalla:2022yfr} for reviews).

Besides the Hubble expansion rate $H(z)$, another fundamental quantity in universes described by the FLRW metric is the distance duality relation (DDR), or Etherington reciprocity theorem, which relates cosmological distances measured through the luminosity and the angular size of astrophysical objects \citep{DDR_original}. The validity of the DDR stems from the metricity of the gravitational theory and the masslessness and number conservation of photons. It therefore holds in any theory of gravity which assumes photons propagate on the null geodesics of the spacetime, regardless of the assumed matter--energy content. It can be  measured directly with ``golden'' observables for which we know both the angular size and luminosity, for example strongly gravitationally lensed SNIa \citep{Renzi:2020bvl}, strongly lensed gravitational waves \citep{Lin:2020vqj}, Type II supernovae \citep{TypeIISNa}, or by combining two or more observables \citep{Liao:2019xug, Hogg:2020ktc, EUCLID:2020syl}.

However, as cosmological distances cannot be directly measured without knowing the value of the Hubble constant, an intrinsic degeneracy exists between $H_0$ and the parameter or parameterisation used to quantify the DDR \citep{Renzi:2020fnx}.  In this paper, we explicitly expose the correlation between $H_0$ and the DDR and provide the first joint measurements of these two quantities by employing a completely model-independent approach. We further show that our constraints can be used to construct a consistency test for cosmological models based on assuming the validity of the DDR.

The paper is organised as follows: in \autoref{sec.II} we review how current cosmological and astrophysical data can be used to build estimators of $H_0$ and the DDR; in \autoref{sec.III} we introduce our machine learning methodology which we use to extract the relevant information from the available data; in \autoref{sec.IV} we show and discuss our results and in \autoref{sec.V} we conclude.

\section{The empirical Etherington--Hubble relation}\label{sec.II}
A spatially flat, homogeneous and isotropic spacetime is described by the Friedmann--Lema\^{i}tre--Robertson--Walker (FRLW) metric,
\begin{equation}
 	\dd s^2 = -\dd t^2 + a^2(t)(\dd\chi^2 + \chi^2 \, \dd\Omega^2),
\end{equation}
where the speed of light $c=1$, $a(t)$ is the cosmic scale factor, $\chi$ is the comoving distance and $\dd\Omega^2=\dd\theta^2 +\sin^2\theta \, \dd\phi^2$ is the metric on a 2-sphere in polar coordinates. By considering the radial propagation of massless particles in this metric, a relation between the evolution of the scale factor and the comoving distance can be derived,
\begin{equation}\label{Eqn.distance_scaling}
\ddf{\chi}{z} = H(z)^{-1},
\end{equation}
where we have used $a(t) = (1+z)^{-1}$ to define the redshift $z$ and introduced the Hubble parameter $H(t) = \dd \ln a/\dd t$ in terms of $ z $, with $H(z=0) = H_0$ being the Hubble constant. This relation is known as Hubble's law.

Cosmological observations cannot directly measure the comoving distance. Instead, they infer either the angular diameter distance, $d_A$, from the angular scale of an object of known size, or the luminosity distance, $d_L$, from the flux of a source of known intrinsic brightness. While they are both related to the comoving distance, these relations are model-dependent. However, for any metric theory of gravity where photons propagate on null geodesics and for which photon number is conserved, the mutual scaling of the luminosity and angular distances with redshift follows the DDR \citep{DDR_original},
\begin{equation}\label{Eqn.DDR}
	\eta(z) \equiv \frac{d_L(z)}{(1+z)^2d_A(z)} = 1,
 \end{equation} 
where the luminosity distance $d_L = (1+z)\chi$ and the angular diameter distance $d_A = \chi/(1+z)$.
Deviations from $\eta(z)=1$ would appear if photons did not propagate on null geodesics (\ie were not massless), if photon number was not conserved (perhaps through a decay of photons into another particle), or if the spacetime was not described by a pseudo-Riemannian manifold\footnote{A theory of gravity is metric if the Christoffel symbols can be expressed through the metric tensor. This implies the manifold to be pseudo-Riemannian.}. Note that the DDR will hold regardless of the form one chooses to parameterise the expansion rate $H(z)$ with, since it is a fundamental relation which stems from the geodesic motion of massless particles \citep{DDR_original}. In particular, it will hold for curved spacetimes where the three-dimensional manifold curvature is different from zero. 

We hazard that it is not a particularly strong assumption to expect $\eta(z)=1$. Nevertheless, it is important to consider what measurements we can make to confirm this expectation.

At first glance, it appears that we can probe deviations from the DDR directly through observations of luminosity and angular distances. However, from the definition of the FLRW metric, it is clear that these distances depend on the reference frame that the observer chooses to describe their spacetime. This means that they are not \textit{true} observables. 

If we have some object of known angular size (a standard ruler), we cannot measure the true (angular) distance to that object. Rather, we measure the ratio of the angular distance to the physical length of the ruler, $\theta^{-1}=d_A(z)/R$, where $R$ is the ruler length and $\theta$ is the measured angular size of the object. If we have a set of standard rulers at different redshifts, we can measure the evolution of $d_A(z)/R$, or alternatively $H(z)R$, using \autoref{Eqn.distance_scaling}. The dependence on the intrinsic ruler length can be eliminated by multiplying the constraints to estimate the combination $d_A(z)H(z)$. 

Similarly, if we have some object of known intrinsic luminosity (a standard or more properly a \textit{standardisable} candle), we cannot measure the true (luminosity) distance to that object. Instead, what we probe is the ratio of two fluxes, or else $ d_L^2/D^2 \sim (H_0d_L)^2 $, where $D$ is the distance we would measure to the source if it were at $z\sim 0$, and we have used  \autoref{Eqn.distance_scaling} to write $D$ in terms of the Hubble constant.

It is therefore convenient to rewrite \autoref{Eqn.DDR} in a form that takes the peculiarities of standard rulers and candles into account,
\begin{equation}\label{Eqn.empirical_DDR}
	\frac{\eta(z)H_0}{H(z)} = \frac{1}{(1+z)^2}\frac{[H_0d_L(z)]^{\rm candle}}{[H(z)d_A(z)]^{\rm ruler}}.
\end{equation}

From \autoref{Eqn.empirical_DDR}, it is therefore possible to derive constraints on both the Hubble constant and the DDR by employing a specific choice for $H(z)$. 

The first possibility is to use the Einstein field equations to calculate the equations of motion associated with the FLRW metric. This yields the Friedmann equation, which relates the cosmic expansion to the matter--energy content of the Universe,
\begin{equation}
	E^2(z)\equiv\frac{H^2(z)}{H^2_0}=\Omega_m (1+z)^3+\Omega_r (1+z)^4+\Omega_{\rm DE}X(z)\,.
\end{equation}
where $E^2(z)$ is the dimensionless expansion rate and $\Omega_i = \rho^0_i/\rho^0_{\rm crit}$ represents the fractional density of the $i^{\rm th}$ component today, $\rho_{\rm crit} = 3H^2/8\pi G$ being the critical energy density. The subscript DE refers to any possible deviation from the standard cosmological model in the form of dark energy or a modification of gravity. The possible time-dependence of this component is expressed through $X(z)$. When $X(z)=1$, we have a cosmological constant and hence $E^2(z)$ describes $\Lambda$CDM. In other words, the parameter $X(z)$ is able to capture any contribution to $E^2(z)$ which does not come from matter or radiation.

This approach is manifestly model-dependent, as a specific choice of $X(z)$ is required to obtain $H(z)$. However, $H(z)$ is a measurable quantity and can be constrained through observations. To show this, we redefine $H(z)$ in the following way,
\begin{equation}
H(z) = \ddf{\ln a}{t} = - \frac{1}{(1+z)} \ddf{z}{t}.
\end{equation}

Finally, we rewrite \autoref{Eqn.empirical_DDR} to eliminate the $H(z)$ dependency,
\begin{align}\label{Eqn.eDDR_final}
    \eta(z) H_0 = \frac{1}{(1+z)^2} \, \frac{[H_0 d_L(z)]^{\rm candle}}{[d_A(z)]^{\rm clock\ +\ ruler}}.
\end{align}
This empirical relation allows us to constrain $H_0$ and $\eta(z)$ at the same time, without needing to explicitly define the form of the cosmological model $E^2(z)$. We call this expression \textit{the Etherington--Hubble relation}.

\section{Reconstructing distances from observational data}\label{sec.III}
In order to fully exploit the Etherington--Hubble relation presented in \autoref{Eqn.eDDR_final}, we must employ a method which does not involve explicitly defining $E^2(z)$. We follow an approach based on Gaussian processes (GPs) to interpolate cosmological data and infer the distance--redshift relation. A Gaussian process is defined as ``a collection of random variables, any finite number of which have a joint Gaussian distribution'' \citep{GPbible}. A GP is therefore a generalisation of a Gaussian probability distribution, but where a probability distribution describes finite-dimensional random variables, a GP  describes the properties of functions, and can be used to reconstruct a function $f(z)$ given the function values $z$, as well as the mean and covariance function of the GP, also known as the kernel. We will discuss the kernel choice when presenting the results of our analysis.

Given a set of observations from cosmological data of some generic function of redshift, $f(z)$, we build a GP interpolation of $f(z)$ tailored to that data. This allows us to obtain a continuous set of probability distribution functions (PDFs) that represent $ f(z) $ at each redshift. From these PDFs we can construct any function, $F(z)=F(f(z))$, by propagating samples (random variates) of the PDFs of $f(z)$ into those of $F(z)$. The propagation $f \rightarrow F $ is  done in a similar way to the evaluation of Markov chain Monte Carlo (MCMC) parameters which are not sampled during the likelihood maximisation \citep{Renzi:2020bvl}. For this reason we refer to this methodology as Gaussian Process Monte Carlo (GPMC), which was first introduced by \cite{Renzi:2020fnx}. We now describe the datasets we use to reconstruct the distances involved in \autoref{Eqn.eDDR_final}.

\subsection{Standard rulers and clocks}

As discussed in the previous section, standard rulers provide two different constraints that can be combined to obtain $d_A(z)H(z)$, while standard clocks provide $H(z)$. Combining rulers and clocks, we can reconstruct the distance--redshift relation in terms of the angular diameter distance. For this purpose, we employ baryon acoustic oscillation (BAO) data from the latest release of the Sloan Digital Sky Survey (SDSS) collaboration as standard rulers, and a collection of observations of $H(z)$ obtained from cosmic chronometers as standard clocks (both datasets are reported in \hyperref[sec.AppendixA]{Appendix A}). To build $d_A(z)$, we start by constructing the $d_A(z)/R$ and $H(z)R$ distributions from the SDSS BAO data, assuming they are Gaussian\footnote{This is a fair assumption considering the symmetry of the bounds on the BAO constraints.}. 
The SDSS collaboration has provided seven such measurements in the range $0.3 \leq z \leq 2.3$, which we use to build $d_A(z)/R$ and $H(z)R$. We then combine them to obtain seven effective measurements of $d_A(z)H(z)$. The second step to obtain $d_A(z)$ is to interpolate the cosmic chronometer data using a GP regression to get the value of $H(z)$ at the same redshift as the BAO data. In this way, we can combine the two datasets into measurements of $d_A(z)$ at the seven BAO redshifts. The combination of the two datasets is done by taking samples of the distributions of $d_A(z)H(z)$ and $H(z)$ and combining them algebraically into a sample of $d_A(z)$. We fix the number of samples at each redshift for all distributions to 10,000, to ensure that the samples provide a fair representation of the PDFs from which they are drawn. We discuss the systematics of the BAO data points in \hyperref[sec.AppendixB]{Appendix B}.

\subsection{Standard candles}
To reconstruct the distance--redshift relation in terms of the luminosity distance we proceed in a similar way to the method used to obtain the angular diameter distance. We use a collection of 1048 B-band observations of the relative magnitudes of Type Ia supernovae, $m_{\mathrm B}(z)$, collectively known as the Pantheon dataset \citep{Scolnic:2017caz} as our standard candles. The first step is to interpolate the Pantheon dataset to infer the value of $m_{\mathrm B}(z)$ at the redshifts of the BAO. The relative supernova magnitude is related to the distance from the supernova itself by the following equation,
\begin{equation}\label{Eqn.SNedistance}
    m_{\mathrm B}(z) - M_{\mathrm B} = 5\log_{10}{d_L(z)} + 25,
\end{equation}
where $M_{\mathrm B}$ is the magnitude any supernova would have if it was at a distance of 10pc from the observer, also known as the supernova absolute magnitude. The value of $M_{\mathrm B}$ can be easily related to the Hubble constant by taking into account that for $z \rightarrow 0 $ the luminosity distance can be determined through the local Hubble law, $d_L(z) = (1+z)\chi  = (1+z)z/H_0$. This allows us to rewrite \autoref{Eqn.SNedistance}, substituting $H_0$ in place of $M_{\mathrm B}$, \ie
\begin{equation}\label{}
    m_{\mathrm B}(z) = 5\log_{10}(H_0d_L(z)) - 5a_{\mathrm B},
\end{equation}
where $a_{\mathrm B}$ is the intercept of the magnitude--redshift relation, approximately given by $\log_{10}z - 0.2 m_{\mathrm B}^0$ with $m_{\mathrm B}^0 = m_{\mathrm B}(z\sim0)$. 

For a generic expansion and $z >0 $, the value of $a_{\mathrm B}$ can be expressed through a cosmographic expansion, as in \cite{Riess:2016jrr},
\begin{multline}
\mathrm{e}^{a_{\mathrm B} + 0.2 m_{\mathrm B}^0} = z \left[1 + \frac{1}{2}(q_0 -1) \,z \right. \\ 
    \left. - \frac{1}{6}(1-q_0-3q_0^2+j_0) \, z^2 + \mathcal{O}(z^3) \vphantom{\frac12}\right].
\end{multline}
In the following we will assume a Gaussian distribution for $a_{\mathrm B}$ with $a_{\mathrm B} = 0.71723 \pm 0.00176$ \citep{Riess:2016jrr}. This value is obtained by fixing $q_0 = -0.55$ and $j_0=1$. However, the dependency of $a_{\mathrm B}$ on the choice of $q_0$ and $j_0$ is negligible given the redshift range in which the fit for $a_{\mathrm B}$ is typically performed ($z \lesssim 0.2$) and current experimental uncertainties \citep{Riess:2016jrr,Camarena:2021jlr}. Combining the prior information on $a_{\mathrm B}$ with the GP reconstruction of $m_{\mathrm B}(z)$ we can finally obtain estimates of the luminosity  distance at the BAO redshifts and use them to constrain the Etherington--Hubble relation. These combinations are again performed in an MCMC-like fashion, algebraically combining the PDF samples at each redshift through \autoref{Eqn.eDDR_final}. 

\section{Results}\label{sec.IV}

\subsection{The GPMC analysis}\label{subsec.IV.A}

\begin{figure}
    \centering
    \includegraphics[width=\columnwidth]{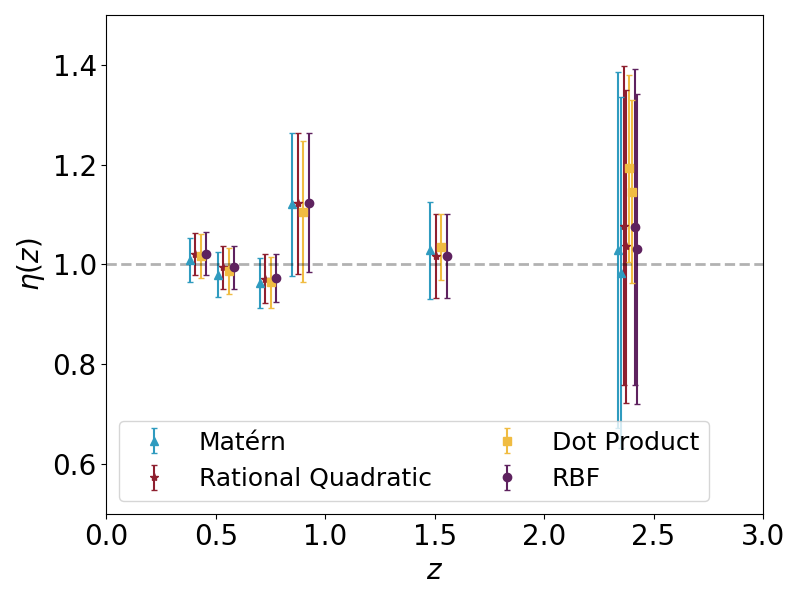}
    \caption{Measurements of the distance duality relation $\eta(z)$ at several redshifts obtained by combining BAO, SNIa and CC data using our GPMC method with four different kernels, and assuming $H_0 = 70$ kms$^{-1}$ Mpc$^{-1}$.}
    \label{fig:eta_of_z_kernels}
\end{figure}

We firstly present the constraints on $\eta(z)$ obtained using the GPMC method described above. We begin by checking the stability of our results against a change in the GP kernel used. In \autoref{fig:eta_of_z_kernels}, we show measurements of $\eta(z)$ when considering four different kernels available in the Python library \texttt{scikit-learn}\footnote{\url{https://scikit-learn.org/}}: the Mat\'ern, Rational Quadratic, Dot Product and Radial Basis Function (RBF) kernels. Note that we introduce a small artificial offset in redshift so that all the points are clearly visible.

From this plot, we can see that the measurement of $\eta(z)$ is extremely consistent across the kernels. For this reason, we choose to focus on the results obtained using the Mat\'ern kernel from now on. This kernel takes the form
\begin{equation}
    k(z, \tilde{z}) = \sigma^2_M \, \frac{2^{1-\nu}}{\Gamma(\nu)} \, \left[\frac{\sqrt{2\nu} \, d(z, \tilde{z})}{\ell}\right]^\nu
     K_\nu \left[ \frac{\sqrt{2\nu} \, d(z, \tilde{z})}{\ell}\right], 
\end{equation}
where $d(z, \tilde{z})$ is the Euclidean distance between the two input points $z$ and $\tilde{z}$, $\Gamma(\nu)$ is the gamma function, $K_\nu$ is a modified Bessel function and $\nu$ controls the shape of the kernel. We fix $\nu = 7/2$ for both the CC and the Pantheon datasets as this yields the lowest $\chi^2$ value \ie the reconstruction which is the best fit to the data. The hyperparameters $\ell$ and $\sigma_M$ represent the length scale over which the reconstructed function varies, and the magnitude of the variations. 

We fix the kernel hyperparameters by minimising the logarithm of a $\chi^2$ likelihood, where the $\chi^2$ is calculated comparing the prediction of the GP interpolation with a test dataset. As test datasets we use the binned Pantheon dataset \citep{Scolnic:2017caz} for the fit to the SNIa data and a collection of six $E(z)$ measurements also obtained from the Pantheon data \citep{Riess:2017lxs} for the fit to the CC data. We perform the minimisation using the \texttt{Cobaya} minimiser \citep{Torrado:2020dgo}, which employs the \texttt{BOBYQA} minimisation algorithm \citep{BOBYQA:3, BOBYQA:1,BOBYQA:2}. 

In \autoref{fig:eta_of_z_kernels}, our results clearly show that the combination of CC, BAO and SNIa data is in agreement with $\eta(z) = 1$ within $1\sigma$ \ie we find no violation of the distance duality relation in our late time data.

Next, we run a full MCMC analysis using the \texttt{Cobaya} package \citep{Torrado:2020dgo}, using the data described above as the likelihood to constrain $H_0$ and $\eta(z)$ at the same time. 
We sample the posterior distribution using the MCMC sampler developed for \texttt{CosmoMC} \citep{Lewis:2002ah,Lewis:2013hha} and tailored for parameter spaces with a speed hierarchy. It also implements the ``fast dragging'' procedure described in \cite{Neal:2005}.

\begin{figure*}
    \centering
    \includegraphics[width=.7\textwidth]{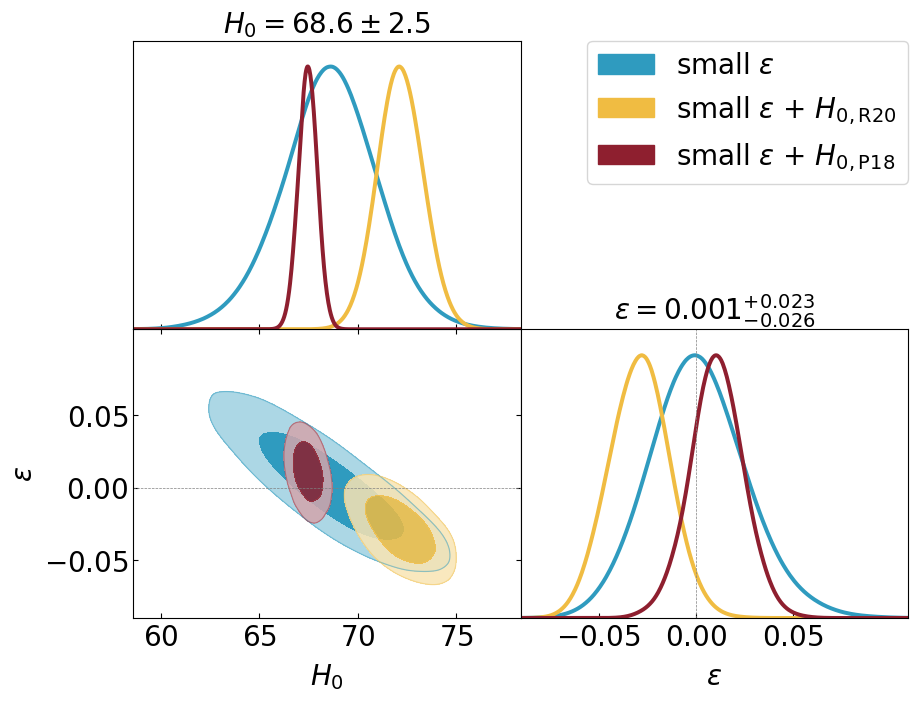}
    \caption{Joint constraints on $H_0$ and $\epsilon$ found using the BAO, SNIa and CC data (blue) and reweighted using a P18 prior (red) and a R20 prior (yellow). The subplot titles give the mean and $1\sigma$ marginalised posterior values of $H_0$ and $\epsilon$ for the unweighted chain (blue).}
    \label{fig:matern_contour}
\end{figure*}

To obtain constraints on $\eta(z)$, we parameterise its evolution in redshift as $\eta(z) = 1+\epsilon z$, meaning that if $\epsilon$ is measured to be zero, $\eta(z)=1$ and we find no violation of the DDR. This parameterisation is motivated by the simple fact that we expect the DDR, if violated, to deviate only slightly from unity, as indicated by current constraints on $\eta(z)$~\citep{Li:2011exa, Ma:2016bjt, Holanda:2015zpz,Holanda:2016msr, Rana:2017sfr, Zhou:2020moc,Holanda:2020fmo,Xu:2020fxj, Bora:2021cjl}. We present the constraints on $H_0$ and $\epsilon$ obtained with our pipeline in \autoref{fig:matern_contour}. Note that we also checked the stability of our results against changes in the parameterisation of $\eta(z)$, finding no differences. Specifically, we found no change in our results using $\eta(z) = 1 + \epsilon z$,  $\eta(z) = 1 + \epsilon (z / (1+z))$,  $\eta(z) = 1 + \epsilon \ln (1+z)$, $\eta(z) = 1 + \epsilon (z / (1+z)^2)$ or $\eta(z) = 1 + \epsilon (z(1+z) / 1+z^2)$. However, in cases where GP results are very sensitive to a change in kernel or the dataset is very large, a genetic algorithms or neural network approach may be more appropriate \citep{Bernardo:2021mfs,Dialektopoulos:2021wde}.

It is clear from \autoref{fig:matern_contour} that there is a degeneracy between $\epsilon$ (or equivalently $\eta(z)$) and $H_0$. We also explicitly show how the imposition of different priors on $H_0$ can alter the measurement of $\epsilon$ obtained from the same data. We reweight the marginalised posterior distributions obtained for $H_0$ using two different Gaussian priors. We use the P18 prior of $H_0 = 67.4 \pm 0.5$ kms$^{-1}$ Mpc$^{-1}$ \citep{Aghanim:2018eyx} (shown in red in \autoref{fig:matern_contour}) and R20 prior of $H_0 = 73.2 \pm 1.3$ kms$^{-1}$ Mpc$^{-1}$ (shown in yellow in \autoref{fig:matern_contour}). This demonstrates how changing the cosmological model (here done implicitly by the choice of prior -- a model-dependent value from the CMB, or a model-independent value from Cepheid-calibrated supernovae) can in principle alter the measurement of the DDR. In other words, the $H_0$ tension can be recast as a tension in measurements made of the DDR.

While the reweighted results are still both consistent with $\eta(z) = 1$ (P18 at 1$\sigma$ and R20 at 2$\sigma$), depending on one's level of tolerance, the $2 \sigma$ discrepancy between the R20 measurement of $\epsilon$ and $\epsilon = 0$ may be taken more or less seriously.

Although its statistical significance is not high enough to draw reliable conclusions, if we are uncomfortable with this level of disagreement between measurements and we trust the DDR, our result can be interpreted as a clue that the local measurements of $H_0$ are biased towards an artificially high value. As many have previously speculated, this could be due to systematic errors in the SNIa calibration \citep{Efstathiou:2021ocp,Camarena:2021jlr,Renzi:2020fnx}. 
Conversely, if we trust the S$H_0$ES measurements, this 2$\sigma$ discrepancy may be seen as a hint for new physics beyond the standard cosmological model at late times, lending weight to the possibility that the solution to the $H_0$ tension could involve the breakdown of the Etherington reciprocity theorem. 

Leaving aside observational systematics, the possible causes of such a violation can be attributed to only a few physical effects. One possibility is that photons are interacting with some particles not in the Standard Model e.g. photon conversion into axion-like particles \citep{Tiwari:2016cps,Buen-Abad:2020zbd,DeBernardis:2006ii,Mirizzi:2006zy,Masaki:2017aea,Masaki:2019ggg,Mukherjee:2018zzg,Avgoustidis:2010ju,DAmico:2015snf}. Another possibility is that photons have some small but non-zero mass. While being a rather exotic avenue, the possibility remains open, even though strong constraints on photon mass have been derived using astrophysical observations \citep{Schaefer:1998zg,Wu:2016brq}. A final possibility is to consider the breakdown of the FLRW metric by non-uniform Hubble flow~\citep{McClure:2007vv,Krishnan:2021dyb,Krishnan:2021jmh,Luongo:2021nqh} or by theories of gravitation which do not respect the metricity ansatz, one example being teleparallel theories \citep{DeAndrade:2000sf,Bahamonde:2021gfp}. Generally speaking, all these possibilities involve modifying the propagation of photons.

Let us close this section with the following remark: the constraint we obtain for $H_0$ is the first ever measurement that does not rely on any cosmological assumptions. All we have assumed is the existence of standardisable observables and that they are all tracers of the same cosmic expansion and therefore related to the same underlying $H(z)$. This means that our constraints on the Hubble constant are completely model-independent. 

\subsection{Changing the cosmological model}
\begin{table*}
\centering
\renewcommand{\arraystretch}{1.5}
\begin{tabular}{l c@{\hspace{1 cm}} c @{\hspace{1 cm}} c @{\hspace{1 cm}} c}
\toprule
 & Inferred $\epsilon$ & Consistency with DDR & Consistency with R20 & Consistency with GPMC \\
Cosmological model & (P18) & ($\epsilon=0$) & ($\epsilon=-0.031\pm 0.016$) & ($\epsilon = 0.001^{+0.023}_{-0.026}$)\\
\hline\hline
		$\Lambda$CDM      &$0.006\pm 0.016$   & $0.4\,\sigma$  & $1.6\,\sigma$ & $0.2\,\sigma$ \\
		$\Lambda$CDM + $M_{\nu}$     &$0.009\pm 0.018$	 & $0.5\,\sigma$  & $1.7\,\sigma$ & $0.2\,\sigma$\\
		$\Lambda$CDM +$ N_{\rm eff}$    &$0.014\pm 0.019$  & $0.7\,\sigma$  & $1.8\,\sigma$ & $0.4\,\sigma$\\
		$\Lambda$CDM + $M_{\nu}$ + $N_{\rm eff}$    &$0.016\pm 0.021$   & $0.8\,\sigma$  & $1.8\,\sigma$ & $0.4\,\sigma$\\
		$\Lambda$CDM + $\Omega_k$  &$0.048^{+0.026}_{-0.029}$	 & $1.6\,\sigma$  & $2.4\,\sigma$ & $1.2\,\sigma$\\
		$\Lambda$CDM + $\Omega_k$ + $M_{\nu}$   &$0.032\pm 0.027$  & $1.2\,\sigma$  & $2.0\,\sigma$ & $0.8\,\sigma$\\
		$\Lambda$CDM + $\Omega_k$ + $N_{\rm eff}$      &$0.044\pm 0.027$   & $1.6\,\sigma$  & $2.4\,\sigma$ & $1.1\,\sigma$\\
		$\Lambda$CDM + $\Omega_k$ + $M_{\nu}$ + $N_{\rm eff}$    &$0.037\pm 0.027$	 & $1.4\,\sigma$  & $2.2\,\sigma$ & $1.0\,\sigma$\\
\bottomrule
    \end{tabular}
    \caption{Measurements of $\epsilon$ considering different cosmological models, and their consistency with the DDR, R20 and the GPMC.}
\label{tab.Results.Models}
\end{table*}

In this section, we now move to describe how the validity of the DDR can be used as a consistency test for extensions to $\Lambda$CDM. Firstly, it is important to note that the full Planck likelihood uses a temperature scaling relation which assumes that photons propagate on null geodesics, \ie
\begin{align}
    T = (1+z)\; T_0,
\end{align}
where $z$ is the redshift and $T_0$ is the observed CMB temperature at $z=0$. This relation is exactly equivalent to rewriting the DDR in terms of the temperature \citep{Avgoustidis:2011aa,Avgoustidis:2015xhk} and therefore implies $\eta(z) = 1$. As the results of a fit to the CMB data can be summarised (in terms of late time observables) with $H_0$, the value obtained for this parameter is forced to be consistent with $\epsilon = 0$. We can see from \autoref{fig:matern_contour} that this is indeed the case for the result obtained using the Planck prior on $H_0$ assuming $\Lambda$CDM expansion. 

As pointed out in \autoref{sec.II}, the validity of the  distance duality relation, \ie $\eta(z)=1$, is a robust assumption to make. We now use the resilience of this relation and the fact that it is implicitly assumed in the CMB fit as a general consistency check for a given cosmological model. 

We firstly estimate the value of $H_0$ using the Planck 2018 observations of the cosmic microwave background temperature anisotropies and polarisation. While neither the results obtained by the Gaussian Process Monte Carlo technique or the local measurement of $H_0$ provided by the S$H_0$ES collaboration rely on the precise form of $E^2(z)$, inferring the value of the present day expansion rate from observations of the early universe (\ie the CMB) necessarily requires a cosmological model.  

We start by considering the standard $\Lambda$CDM model based on the usual six free parameters and then proceed by including different combinations of the additional parameters, such as the total neutrino mass $M_{\nu}\equiv \sum m_{\nu}$,  the effective number of relativistic species $N_{\rm eff}$ and the spatial curvature parameter $\Omega_k$. Indeed, it is well known that the constraints on $H_0$ can be significantly changed by including additional degrees of freedom in the sample~\citep{DiValentino:2016hlg,DiValentino:2019dzu,DiValentino:2019qzk,DiValentino:2020hov,DiValentino:2021izs,Perivolaropoulos:2021jda,Anchordoqui:2021gji}. 

For instance, as robustly indicated by oscillation experiments~\citep{deSalas:2020pgw,deSalas:2018bym}, neutrinos should be regarded as massive particles, and cosmology provides a powerful (albeit indirect) means to constrain their mass~\citep{deSalas:2018bym,Hagstotz:2020ukm,Vagnozzi:2019utt,Vagnozzi:2018pwo,Vagnozzi:2018jhn,Vagnozzi:2017ovm,Giusarma:2016phn,Bond:1980ha,Capozzi:2021fjo,DiValentino:2021hoh,Xu:2020fyg,Green:2021xzn}. The strong correlation between the expansion rate and the total neutrino mass can alter the final constraint on $H_0$.

Similarly, by testing departures from the reference value $N_{\rm eff}=3.046$~\citep{Mangano:2005cc,Akita:2020szl,Froustey:2020mcq,Bennett:2020zkv,Schwarz:2003du}, one can probe and constrain several extended models both of cosmology and particle physics that predict extra dark radiation in the early Universe, including the cases of additional neutrino species and hot relics beyond the standard model of elementary particles~\citep{Melchiorri:2007cd,DiValentino:2015zta,DiValentino:2015wba,Archidiacono:2015mda,DEramo:2018vss,Giare:2020vzo,Giare:2021cqr,Green:2021hjh,DEramo:2021usm,DePorzio:2020wcz,DEramo:2021lgb,DEramo:2021psx}. Higher values of the effective number of relativistic species can lead to smaller values of the sound horizon at recombination, resulting in a preference for higher values of $H_0$.

Finally, the last Planck Collaboration data release~\citep{Akrami:2018vks, Aghanim:2019ame,Aghanim:2018eyx} confirmed the presence of an enhanced lensing amplitude in CMB power spectra, larger than what is expected in the standard $\Lambda$CDM model. While such a preference could be a manifestation of some unaccounted-for systematics in the CMB data \citep{Efstathiou:2020wem,Vagnozzi:2020zrh,Vagnozzi:2020dfn}, as argued by \cite{DiValentino:2019qzk,Handley:2019tkm}, the possibility of a closed Universe ($\Omega_k < 0$) can provide another valid physical explanation for this effect\footnote{Other possible explanations to this effect involve for instance modified gravity theories~\citep{Raveri:2021dbu,Aoki:2020oqc}}. Despite the fact that inflationary theory naturally predicts a spatially flat geometry, it is nevertheless possible to construct inflationary models with positive curvature ~\citep{Linde:1995xm,Linde:2003hc,Ratra:2017ezv,Bonga:2016iuf,Handley:2019anl,Bonga:2016cje,Ooba:2017ukj,Ellis:2001ym,Uzan:2003nk,Unger:2018oqo,Gordon:2020gel,Sloan:2019jyl,Motaharfar:2021gwi}. When the assumption of flatness is relaxed, the analysis of Planck CMB temperature and polarisation (TT TE EE) data show a preference for a closed Universe at more than 95\% confidence level (CL hereafter) \citep{Aghanim:2018eyx,DiValentino:2019qzk,Handley:2019tkm,DiValentino:2020hov, DiValentino:2020srs, Forconi:2021que,Zuckerman:2021kgm}. In such models, the Hubble tension is substantially increased and a large discordance arises in most of the local cosmological observables compared to what is observed by other means, including BAO\footnote{It is worth noting that including in the analysis the lensing spectrum as measured by the Planck Collaboration and the BAO data we recover the preference for a flat Universe~\citep{Akrami:2018vks,Forconi:2021que}. However these results should be considered with caution because local measurements are in strong disagreement with Planck when the curvature parameter is free to vary~\citep{DiValentino:2019qzk,Handley:2019tkm,DiValentino:2020hov} } ~\citep{DiValentino:2019qzk,Handley:2019tkm,DiValentino:2020hov} . Since the DDR is preserved in a closed Universe, we can maintain an agnostic perspective on the spatial geometry and examine several extended non-flat cosmologies to study if the aforementioned tension can be recast in terms of a violation of the DDR.

\begin{figure}
\centering
\includegraphics[width=\columnwidth]{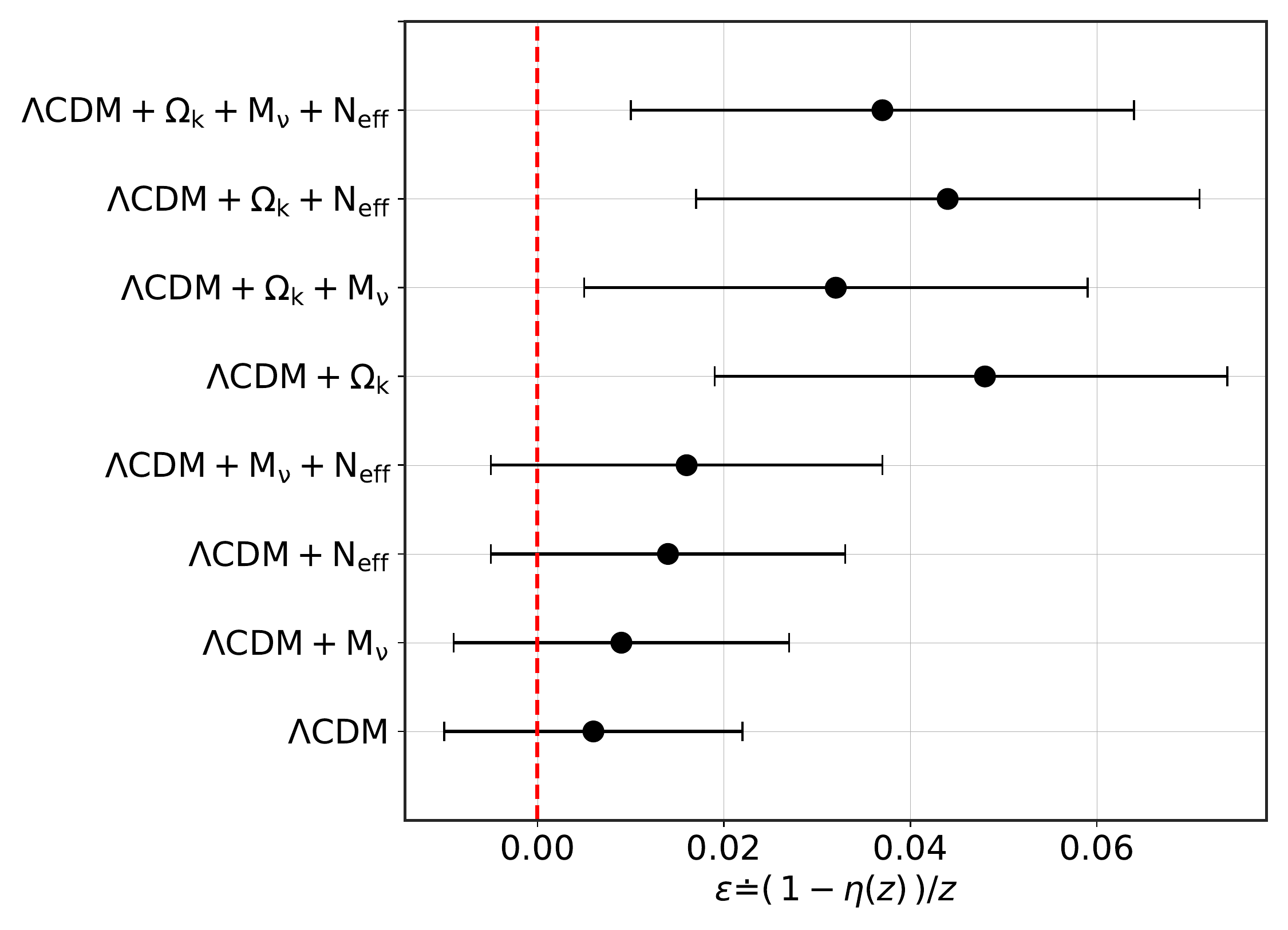}
\caption{Values of $\epsilon$ and their $1\sigma$ uncertainties inferred in different cosmological models when adopting the Planck 2018 values of $H_0$. 
The red dashed line represents the value of $\epsilon$ for which $\eta(z) = 1$.}
\label{fig:WhiskerPlot}
\end{figure}

For each model analysed, we perform a Markov chain Monte Carlo parameter estimation using the Planck 2018 likelihood, in each case deriving a posterior distribution function for $H_0$. We adopt these posterior distributions in the likelihood for the distance duality relation and therefore infer the value of $\epsilon$ (defined as before by the relation $\eta(z)=1+\epsilon z$) in all the different models. This allows us to use the DDR as a consistency check for the different models and to study whether the tensions between early and late time estimations of $H_0$ are also reflected in $\epsilon$. We present our results for the different models in \autoref{tab.Results.Models} and we summarise them in \autoref{fig:WhiskerPlot}.

As already discussed in \autoref{subsec.IV.A}, considering the value of $H_0$ obtained by the Planck Collaboration within the standard $\Lambda$CDM model, we infer $\epsilon=0.006\pm 0.016$ at 68\% CL, consistent with the DDR at $0.4\,\sigma$. This value is also in perfect agreement ($0.2\,\sigma$)  with the result obtained following the GPMC technique. Interestingly, the tension between the Planck estimation of $H_0$ and the local measurement of the S$H_0$ES collaboration is reflected in a difference in the respective values of $\epsilon$ at the level of $1.6\sigma$.

We note that the inclusion of the total neutrino mass and the number of relativistic degrees of freedom as additional parameters of our cosmological model does not significantly change our results. In particular, the value of $\epsilon$ inferred in these models always remains consistent with the DDR and the GPMC estimation to within one standard deviation, while the difference with respect to the value of $\epsilon$ inferred when adopting the value of $H_0$ provided by R20 remains at the level of $\sim 1.7\,\sigma$.

Conversely, the addition of spatial curvature significantly increases the differences among the results discussed so far. In particular, adopting the Planck bound on $H_0$ derived within the $\Lambda$CDM + $\Omega_k$ model, we obtain $\epsilon=0.048^{+0.026}_{-0.029}$ at 68\% CL, revealing a discrepancy with respect to the DDR prediction at the level of $1.6\sigma$. Furthermore, the tension with respect to the value of $\epsilon$ inferred when adopting the R20 prior on $H_0$ is increased to the level of $2.4\sigma$. These results are not robust enough to conclude that $\Lambda$CDM + $\Omega_k$ is in disagreement with the DDR, but it is still worth noting the shift towards higher positive values of $\epsilon$ when considering a curved Universe. This shift is due to the Planck preference for lower values of $H_0$ when curvature is varied in the MCMC sampling, and to the strong anti-correlation between $H_0$ and $\epsilon$ -- see also \autoref{fig:matern_contour}.

The tensions between the different datasets are slightly reduced when considering the total neutrino mass and/or the effective number of relativistic degrees of freedom in combination with curvature. In these cases, the less stringent Planck bounds on $H_0$ allow for smaller mean values of $\epsilon$ to be recovered. In any case, we observe a deviation from the DDR at more than $1\sigma$, and the differences with respect to the value of $\epsilon$ extrapolated when reweighting the posterior distributions using the R20 likelihood for $H_0$ always lie in the range between 2 and 2.4 $\sigma$.

\section{Discussion}

During the final stages of this work, a new S$H_0$ES measurement of $H_0$ was released, along with the addition of approximately 700 new SNIa to the Pantheon catalogue \citep{Scolnic:2021amr}. The new measurement is $H_0 = 73.04 \pm 1.04$ kms$^{-1}$ Mpc$^{-1}$ \citep{Riess:2021jrx}. We found that the smaller error bars on this measurement with respect to R20 slightly alters the reweighted constraint on $\epsilon$, leading to a roughly $2.5\sigma$ separation from $\eta(z)=1$. Furthermore, the improved precision of the new S$H_0$ES measurement only serves to strengthen our conclusions about this discrepancy. A full analysis using our GPMC pipeline will be possible when the updated Pantheon catalogue and likelihood becomes available.

It is also worth noting that in using the value of $a_B$ of \cite{Riess:2016jrr}, we are implicitly assuming that the FRLW metric is the correct representation of the expansion history at all redshifts in order to perform the cosmographic expansion of nearby distances. While this is a fair assumption, recent analyses of the Pantheon catalogue have outlined a discrepancy between the high and low redshift bins of the datasets \citep{Dainotti:2021pqg,Kazantzidis:2020xta,DiValentino:2020kha}. If confirmed by other independent observations, this could mean that the cosmographic expansion used to fit the magnitude-Hubble intercept is incorrect. We stress however that the latest release of the S$H_0$ES collaboration \citep{Riess:2021jrx} has reanalysed such assumptions, finding no inconsistency with the assumptions of \cite{Riess:2016jrr}.

Although not the focus of this work, an important avenue of recent interest is the possibility that the $H_0$ tension could be resolved with ultra late-time effects, such as in the context of modified gravity \citep{Perivolaropoulos:2021jda}, by assuming that $H_0$ varies with redshift \citep{Dainotti:2021pqg, Dainotti2022} or by a steep ``phase transition'' in the expansion history or in the SNIa absolute magnitude \citep{Marra:2021fvf,Camarena:2021jlr}. Many of these theories warrant deeper investigation. However, including these theories in a well-defined set-up which can be used to constrain the cosmic history, such as the pipeline which we have presented in this paper, is far from trivial. As we discuss in section \ref{sec.IV}, such models would alter the scaling of the CMB temperature-redshift relation possibly leading to spectral distortions in the CMB black-body spectrum \cite{Chluba:2014wda}. 

Furthermore, the analysis we perform to compare our constraints with the results from Planck would be inconsistent with these models, since the Boltzmann solver used to analyse Planck data assumes a standard temperature-scaling relation \cite{Ivanov:2020mfr}. We also note that by choosing the relation between the magnitude of SNIa and $H_0$, we are implicitly assuming an FRLW metric which again assumes $H_0$ is a constant.

We therefore leave a thorough investigation of such late-time modifications of the expansion history for a future work, since, in a nutshell, the aim of this current work was to use the resilience of the Etherington–Hubble relation as a consistency check for beyond $\Lambda$CDM models, rather than explicitly investigating the $H_0$ tension or an apparent $H_0$ evolution. 

\section{Conclusion}\label{sec.V}
The Etherington reciprocity theorem relates the mutual scaling of the luminosity and angular distances with redshift in any metric theory of gravity where photons propagate on null geodesics and for which photon number is conserved. Given its resilience, the Etherington reciprocity theorem, also known as the distance duality relation, can be regarded as a robust property of any Universe described by an FLRW line element where fundamental interactions obey the Standard Model of particle physics. In this work, we explicitly investigated the degeneracy of the DDR with the Hubble rate, constructing a consistency test methodology for any cosmological model based only on the assumption of the validity of the DDR.

We firstly followed a GPMC procedure to measure the DDR function $\eta(z)$ using a dataset of SNIa, BAO and CC, finding that the data is in agreement with $\eta(z) = 1$ at the level of $1 \sigma$; in other words we find no violation of the DDR in our collection of late time data. We then obtained joint constraints on $H_0$ and the DDR parameter $\epsilon$, revealing the degeneracy between these two quantities. By imposing two different priors on $H_0$, one from P18 and one from R20, we showed how the constraint on $\epsilon$ changes for these two datasets.

In particular, we found the R20 prior to be in a mild statistical discrepancy ($2\sigma$) with the constraints from late time data. While our results are not accurate enough to provide a statistically significant answer to the Hubble tension, they support the argument that the present discrepancy in the value of the Hubble constant may simply arise from unaccounted-for systematics, in line with the findings of other recent studies \citep{Efstathiou:2021ocp,Renzi:2020fnx}. However, it is also interesting to consider the idea that, if the discrepancy is not due to systematics, the solution to the $H_0$ tension could involve the breakdown of the Etherington reciprocity theorem. As we explained in section \ref{sec.IV}, this may be a signal for the need to build a theory of gravity that is not constructed using a pseudo-Riemannian manifold or to consider new physics beyond the Standard Model of elementary interactions, involving mechanisms of photon conversion into other particles or even more exotic scenarios. When considering P18 data and a $\Lambda$CDM cosmology, we found instead an almost perfect agreement with the DDR validity, as expected by the fact that the fit to CMB data assumes the standard temperature scaling relation.

In the second part of our results, we used the finding that the DDR is valid and robust as a consistency check for different cosmological models. When testing the model of $\Lambda$CDM plus a varying total neutrino mass and number of relativistic degrees of freedom, we found that the value of $\epsilon$ agrees with the DDR at the level of $1\sigma$. However, when considering a model in which the spatial curvature is free to vary, we found a value of $\epsilon$ which disagrees with the validity of the DDR at around $1.6\sigma$. This implies that, if we trust that the DDR is valid, a cosmological model of $\Lambda$CDM + $\Omega_k$ is in turn disfavoured at the level of $1.6\sigma$. Given that the direction of this discrepancy is opposite (in terms of $\epsilon$) to that of the R20 prior we conclude that it is a manifestation of the well-known discrepancy between Planck and late time data when a non-flat spacetime is considered ~\citep{DiValentino:2019qzk,Handley:2019tkm,DiValentino:2020hov}.

In conclusion, we have demonstrated how the DDR can be used as a powerful consistency test for beyond $\Lambda$CDM models and, generally speaking, for cosmological measurements involving the Hubble constant, an endeavour which is becoming ever more pressing as we move relentlessly forward into the epoch of ultra-precision cosmology. 


\section*{Acknowledgements}
We thank Eleonora Di Valentino, Alessandro Melchiorri and Matteo Martinelli for useful comments and suggestions during the preparation of this work. FR acknowledges support from the NWO and the Dutch Ministry of Education, Culture and Science (OCW) (through NWO VIDI Grant No.2019/ENW/00678104 and from the D-ITP consortium). NBH is supported by a postdoctoral position funded through two ``la Caixa'' Foundation fellowships (ID 100010434), with fellowship codes LCF/BQ/PI19/11690015 and LCF/BQ/PI19/11690018. WG is supported by ``Theoretical astroparticle Physics'' (TASP), iniziativa specifica INFN.

\section*{Data availability}
The data underlying this article will be shared on reasonable request to the corresponding author.

\bibliographystyle{mnras}
\bibliography{MNRAS} 

\begin{thebibliography}{}
\makeatletter
\relax
\def\mn@urlcharsother{\let\do\@makeother \do\$\do\&\do\#\do\^\do\_\do\%\do\~}
\def\mn@doi{\begingroup\mn@urlcharsother \@ifnextchar [ {\mn@doi@}
  {\mn@doi@[]}}
\def\mn@doi@[#1]#2{\def\@tempa{#1}\ifx\@tempa\@empty \href
  {http://dx.doi.org/#2} {doi:#2}\else \href {http://dx.doi.org/#2} {#1}\fi
  \endgroup}
\def\mn@eprint#1#2{\mn@eprint@#1:#2::\@nil}
\def\mn@eprint@arXiv#1{\href {http://arxiv.org/abs/#1} {{\tt arXiv:#1}}}
\def\mn@eprint@dblp#1{\href {http://dblp.uni-trier.de/rec/bibtex/#1.xml}
  {dblp:#1}}
\def\mn@eprint@#1:#2:#3:#4\@nil{\def\@tempa {#1}\def\@tempb {#2}\def\@tempc
  {#3}\ifx \@tempc \@empty \let \@tempc \@tempb \let \@tempb \@tempa \fi \ifx
  \@tempb \@empty \def\@tempb {arXiv}\fi \@ifundefined
  {mn@eprint@\@tempb}{\@tempb:\@tempc}{\expandafter \expandafter \csname
  mn@eprint@\@tempb\endcsname \expandafter{\@tempc}}}

\bibitem[\protect\citeauthoryear{Abdalla et~al.}{Abdalla
  et~al.}{2022}]{Abdalla:2022yfr}
Abdalla E.,  et~al., 2022, in {2022 Snowmass Summer Study}.  (\mn@eprint
  {arXiv} {2203.06142})

\bibitem[\protect\citeauthoryear{Aghanim et~al.}{Aghanim
  et~al.}{2020a}]{Akrami:2018vks}
Aghanim N.,  et~al., 2020a, \mn@doi [A\&A.] {10.1051/0004-6361/201833880}, 641,
  A1

\bibitem[\protect\citeauthoryear{Aghanim et~al.}{Aghanim
  et~al.}{2020b}]{Aghanim:2019ame}
Aghanim N.,  et~al., 2020b, \mn@doi [Astron. Astrophys.]
  {10.1051/0004-6361/201936386}, 641, A5

\bibitem[\protect\citeauthoryear{Aghanim et~al.}{Aghanim
  et~al.}{2020c}]{Aghanim:2018eyx}
Aghanim N.,  et~al., 2020c, \mn@doi [Astron. Astrophys.]
  {10.1051/0004-6361/201833910}, 641, A6

\bibitem[\protect\citeauthoryear{Akita \& Yamaguchi}{Akita \&
  Yamaguchi}{2020}]{Akita:2020szl}
Akita K.,  Yamaguchi M.,  2020, \mn@doi [JCAP] {10.1088/1475-7516/2020/08/012},
  08, 012

\bibitem[\protect\citeauthoryear{Alam et~al.}{Alam et~al.}{2017}]{Alam:2016hwk}
Alam S.,  et~al., 2017, \mn@doi [MNRAS] {10.1093/mnras/stx721}, 470, 2617

\bibitem[\protect\citeauthoryear{Anchordoqui, Di~Valentino, Pan  \&
  Yang}{Anchordoqui et~al.}{2021}]{Anchordoqui:2021gji}
Anchordoqui L.~A.,  Di~Valentino E.,  Pan S.,   Yang W.,  2021, \mn@doi [JHEAp]
  {10.1016/j.jheap.2021.08.001}, 32, 121

\bibitem[\protect\citeauthoryear{Aoki, De~Felice, Mukohyama, Noui, Oliosi  \&
  Pookkillath}{Aoki et~al.}{2020}]{Aoki:2020oqc}
Aoki K.,  De~Felice A.,  Mukohyama S.,  Noui K.,  Oliosi M.,   Pookkillath
  M.~C.,  2020, \mn@doi [Eur. Phys. J. C] {10.1140/epjc/s10052-020-8291-1}, 80,
  708

\bibitem[\protect\citeauthoryear{Archidiacono, Basse, Hamann, Hannestad,
  Raffelt  \& Wong}{Archidiacono et~al.}{2015}]{Archidiacono:2015mda}
Archidiacono M.,  Basse T.,  Hamann J.,  Hannestad S.,  Raffelt G.,   Wong Y.
  Y.~Y.,  2015, \mn@doi [JCAP] {10.1088/1475-7516/2015/05/050}, 05, 050

\bibitem[\protect\citeauthoryear{Arias-Arag\'on, D'eramo, Ferreira, Merlo  \&
  Notari}{Arias-Arag\'on et~al.}{2020}]{Arias-Aragon:2020qtn}
Arias-Arag\'on F.,  D'eramo F.,  Ferreira R.~Z.,  Merlo L.,   Notari A.,  2020,
  \mn@doi [JCAP] {10.1088/1475-7516/2020/11/025}, 11, 025

\bibitem[\protect\citeauthoryear{Arias-Arag\'on, D'Eramo, Ferreira, Merlo  \&
  Notari}{Arias-Arag\'on et~al.}{2021}]{Arias-Aragon:2020shv}
Arias-Arag\'on F.,  D'Eramo F.,  Ferreira R.~Z.,  Merlo L.,   Notari A.,  2021,
  \mn@doi [JCAP] {10.1088/1475-7516/2021/03/090}, 03, 090

\bibitem[\protect\citeauthoryear{Aubourg et~al.}{Aubourg
  et~al.}{2015}]{Aubourg:2014yra}
Aubourg E.,  et~al., 2015, \mn@doi [Phys. Rev. D] {10.1103/PhysRevD.92.123516},
  92, 123516

\bibitem[\protect\citeauthoryear{Avgoustidis, Burrage, Redondo, Verde  \&
  Jimenez}{Avgoustidis et~al.}{2010}]{Avgoustidis:2010ju}
Avgoustidis A.,  Burrage C.,  Redondo J.,  Verde L.,   Jimenez R.,  2010,
  \mn@doi [JCAP] {10.1088/1475-7516/2010/10/024}, 10, 024

\bibitem[\protect\citeauthoryear{Avgoustidis, Luzzi, Martins  \&
  Monteiro}{Avgoustidis et~al.}{2012}]{Avgoustidis:2011aa}
Avgoustidis A.,  Luzzi G.,  Martins C. J. A.~P.,   Monteiro A. M. R. V.~L.,
  2012, \mn@doi [JCAP] {10.1088/1475-7516/2012/02/013}, 02, 013

\bibitem[\protect\citeauthoryear{Avgoustidis, G\'enova-Santos, Luzzi  \&
  Martins}{Avgoustidis et~al.}{2016}]{Avgoustidis:2015xhk}
Avgoustidis A.,  G\'enova-Santos R.~T.,  Luzzi G.,   Martins C. J. A.~P.,
  2016, \mn@doi [Phys. Rev. D] {10.1103/PhysRevD.93.043521}, 93, 043521

\bibitem[\protect\citeauthoryear{Bahamonde et~al.,}{Bahamonde
  et~al.}{2021}]{Bahamonde:2021gfp}
Bahamonde S.,  et~al., 2021, {Teleparallel Gravity: From Theory to Cosmology}
  (\mn@eprint {arXiv} {2106.13793})

\bibitem[\protect\citeauthoryear{Bautista et~al.}{Bautista
  et~al.}{2020}]{Bautista:2020ahg}
Bautista J.~E.,  et~al., 2020, \mn@doi [MNRAS] {10.1093/mnras/staa2800}, 500,
  736

\bibitem[\protect\citeauthoryear{Bennett et~al.,}{Bennett
  et~al.}{1996}]{Bennett:1996ce}
Bennett C.~L.,  et~al., 1996, \mn@doi [Astrophys. J. Lett.] {10.1086/310075},
  464, L1

\bibitem[\protect\citeauthoryear{Bennett, Buldgen, de Salas, Drewes, Gariazzo,
  Pastor  \& Wong}{Bennett et~al.}{2020}]{Bennett:2020zkv}
Bennett J.~J.,  Buldgen G.,  de Salas P.~F.,  Drewes M.,  Gariazzo S.,  Pastor
  S.,   Wong Y. Y.~Y.,  2020, arXiv:2012.02726

\bibitem[\protect\citeauthoryear{Bernardo \& Levi~Said}{Bernardo \&
  Levi~Said}{2021}]{Bernardo:2021mfs}
Bernardo R.~C.,  Levi~Said J.,  2021, \mn@doi [JCAP]
  {10.1088/1475-7516/2021/08/027}, 08, 027

\bibitem[\protect\citeauthoryear{Blomqvist et~al.}{Blomqvist
  et~al.}{2019}]{Blomqvist:2019rah}
Blomqvist M.,  et~al., 2019, \mn@doi [Astron. Astrophys.]
  {10.1051/0004-6361/201935641}, 629, A86

\bibitem[\protect\citeauthoryear{Bond, Efstathiou  \& Silk}{Bond
  et~al.}{1980}]{Bond:1980ha}
Bond J.~R.,  Efstathiou G.,   Silk J.,  1980, \mn@doi [Phys. Rev. Lett.]
  {10.1103/PhysRevLett.45.1980}, 45, 1980

\bibitem[\protect\citeauthoryear{Bonga, Gupt  \& Yokomizo}{Bonga
  et~al.}{2016}]{Bonga:2016iuf}
Bonga B.,  Gupt B.,   Yokomizo N.,  2016, \mn@doi [JCAP]
  {10.1088/1475-7516/2016/10/031}, 10, 031

\bibitem[\protect\citeauthoryear{Bonga, Gupt  \& Yokomizo}{Bonga
  et~al.}{2017}]{Bonga:2016cje}
Bonga B.,  Gupt B.,   Yokomizo N.,  2017, \mn@doi [JCAP]
  {10.1088/1475-7516/2017/05/021}, 05, 021

\bibitem[\protect\citeauthoryear{Bonilla, Kumar  \& Nunes}{Bonilla
  et~al.}{2021}]{Bonilla:2020wbn}
Bonilla A.,  Kumar S.,   Nunes R.~C.,  2021, \mn@doi [Eur. Phys. J. C]
  {10.1140/epjc/s10052-021-08925-z}, 81, 127

\bibitem[\protect\citeauthoryear{Bora \& Desai}{Bora \&
  Desai}{2021}]{Bora:2021cjl}
Bora K.,  Desai S.,  2021, \mn@doi [JCAP] {10.1088/1475-7516/2021/06/052}, 06,
  052

\bibitem[\protect\citeauthoryear{Buen-Abad, Fan  \& Sun}{Buen-Abad
  et~al.}{2020}]{Buen-Abad:2020zbd}
Buen-Abad M.~A.,  Fan J.,   Sun C.,  2020, {Constraints on Axions from Cosmic
  Distance Measurements} (\mn@eprint {arXiv} {2011.05993})

\bibitem[\protect\citeauthoryear{Camarena \& Marra}{Camarena \&
  Marra}{2021}]{Camarena:2021jlr}
Camarena D.,  Marra V.,  2021, \mn@doi [Mon. Not. Roy. Astron. Soc.]
  {10.1093/mnras/stab1200}, 504, 5164

\bibitem[\protect\citeauthoryear{Capozzi, Di~Valentino, Lisi, Marrone,
  Melchiorri  \& Palazzo}{Capozzi et~al.}{2021}]{Capozzi:2021fjo}
Capozzi F.,  Di~Valentino E.,  Lisi E.,  Marrone A.,  Melchiorri A.,   Palazzo
  A.,  2021, \mn@doi [Phys. Rev. D] {10.1103/PhysRevD.104.083031}, 104, 083031

\bibitem[\protect\citeauthoryear{{Cartis}, {Fiala}, {Marteau}  \&
  {Roberts}}{{Cartis} et~al.}{2018a}]{BOBYQA:1}
{Cartis} C.,  {Fiala} J.,  {Marteau} B.,   {Roberts} L.,  2018a, arXiv
  e-prints, \href {https://ui.adsabs.harvard.edu/abs/2018arXiv180400154C} {p.
  arXiv:1804.00154}

\bibitem[\protect\citeauthoryear{{Cartis}, {Roberts}  \&
  {Sheridan-Methven}}{{Cartis} et~al.}{2018b}]{BOBYQA:2}
{Cartis} C.,  {Roberts} L.,   {Sheridan-Methven} O.,  2018b, arXiv e-prints,
  \href {https://ui.adsabs.harvard.edu/abs/2018arXiv181211343C} {p.
  arXiv:1812.11343}

\bibitem[\protect\citeauthoryear{Chluba}{Chluba}{2014}]{Chluba:2014wda}
Chluba J.,  2014, \mn@doi [Mon. Not. Roy. Astron. Soc.]
  {10.1093/mnras/stu1260}, 443, 1881

\bibitem[\protect\citeauthoryear{D'Amico \& Kaloper}{D'Amico \&
  Kaloper}{2015}]{DAmico:2015snf}
D'Amico G.,  Kaloper N.,  2015, \mn@doi [Phys. Rev. D]
  {10.1103/PhysRevD.91.085015}, 91, 085015

\bibitem[\protect\citeauthoryear{D'Eramo \& Yun}{D'Eramo \&
  Yun}{2021}]{DEramo:2021usm}
D'Eramo F.,  Yun S.,  2021, {Flavor Violating Axions in the Early Universe}
  (\mn@eprint {arXiv} {2111.12108})

\bibitem[\protect\citeauthoryear{D'Eramo, Ferreira, Notari  \& Bernal}{D'Eramo
  et~al.}{2018}]{DEramo:2018vss}
D'Eramo F.,  Ferreira R.~Z.,  Notari A.,   Bernal J.~L.,  2018, \mn@doi [JCAP]
  {10.1088/1475-7516/2018/11/014}, 11, 014

\bibitem[\protect\citeauthoryear{D'Eramo, Hajkarim  \& Yun}{D'Eramo
  et~al.}{2021a}]{DEramo:2021psx}
D'Eramo F.,  Hajkarim F.,   Yun S.,  2021a, {Thermal axion production at low
  temperatures: a smooth treatment of the QCD phase transition} (\mn@eprint
  {arXiv} {2108.04259})

\bibitem[\protect\citeauthoryear{D'Eramo, Hajkarim  \& Yun}{D'Eramo
  et~al.}{2021b}]{DEramo:2021lgb}
D'Eramo F.,  Hajkarim F.,   Yun S.,  2021b, \mn@doi [JHEP]
  {10.1007/JHEP10(2021)224}, 10, 224

\bibitem[\protect\citeauthoryear{Dainotti, De~Simone, Schiavone, Montani,
  Rinaldi  \& Lambiase}{Dainotti et~al.}{2021}]{Dainotti:2021pqg}
Dainotti M.~G.,  De~Simone B.,  Schiavone T.,  Montani G.,  Rinaldi E.,
  Lambiase G.,  2021, \mn@doi [Astrophys. J.] {10.3847/1538-4357/abeb73}, 912,
  150

\bibitem[\protect\citeauthoryear{{Dainotti}, {De Simone}, {Schiavone},
  {Montani}, {Rinaldi}, {Lambiase}, {Bogdan}  \& {Ugale}}{{Dainotti}
  et~al.}{2022}]{Dainotti2022}
{Dainotti} M.~G.,  {De Simone} B.,  {Schiavone} T.,  {Montani} G.,  {Rinaldi}
  E.,  {Lambiase} G.,  {Bogdan} M.,   {Ugale} S.,  2022, arXiv e-prints, \href
  {https://ui.adsabs.harvard.edu/abs/2022arXiv220109848D} {p. arXiv:2201.09848}

\bibitem[\protect\citeauthoryear{De~Andrade, Guillen  \& Pereira}{De~Andrade
  et~al.}{2000}]{DeAndrade:2000sf}
De~Andrade V.~C.,  Guillen L. C.~T.,   Pereira J.~G.,  2000, in {9th Marcel
  Grossmann Meeting on Recent Developments in Theoretical and Experimental
  General Relativity, Gravitation and Relativistic Field Theories (MG 9)}.
  (\mn@eprint {arXiv} {gr-qc/0011087})

\bibitem[\protect\citeauthoryear{De~Bernardis, Giusarma  \&
  Melchiorri}{De~Bernardis et~al.}{2006}]{DeBernardis:2006ii}
De~Bernardis F.,  Giusarma E.,   Melchiorri A.,  2006, \mn@doi [Int. J. Mod.
  Phys. D] {10.1142/S0218271806008486}, 15, 759

\bibitem[\protect\citeauthoryear{De~Salas, Gariazzo, Mena, Ternes  \&
  T\'ortola}{De~Salas et~al.}{2018}]{deSalas:2018bym}
De~Salas P.~F.,  Gariazzo S.,  Mena O.,  Ternes C.~A.,   T\'ortola M.,  2018,
  \mn@doi [Front. Astron. Space Sci.] {10.3389/fspas.2018.00036}, 5, 36

\bibitem[\protect\citeauthoryear{DePorzio, Xu, Mu\~noz  \& Dvorkin}{DePorzio
  et~al.}{2021}]{DePorzio:2020wcz}
DePorzio N.,  Xu W.~L.,  Mu\~noz J.~B.,   Dvorkin C.,  2021, \mn@doi [Phys.
  Rev. D] {10.1103/PhysRevD.103.023504}, 103, 023504

\bibitem[\protect\citeauthoryear{Di~Valentino, Gariazzo, Giusarma  \&
  Mena}{Di~Valentino et~al.}{2015a}]{DiValentino:2015zta}
Di~Valentino E.,  Gariazzo S.,  Giusarma E.,   Mena O.,  2015a, \mn@doi [Phys.
  Rev. D] {10.1103/PhysRevD.91.123505}, 91, 123505

\bibitem[\protect\citeauthoryear{Di~Valentino, Melchiorri  \&
  Silk}{Di~Valentino et~al.}{2015b}]{DiValentino:2015ola}
Di~Valentino E.,  Melchiorri A.,   Silk J.,  2015b, \mn@doi [Phys. Rev. D]
  {10.1103/PhysRevD.92.121302}, 92, 121302

\bibitem[\protect\citeauthoryear{Di~Valentino, Giusarma, Lattanzi, Mena,
  Melchiorri  \& Silk}{Di~Valentino et~al.}{2016a}]{DiValentino:2015wba}
Di~Valentino E.,  Giusarma E.,  Lattanzi M.,  Mena O.,  Melchiorri A.,   Silk
  J.,  2016a, \mn@doi [Phys. Lett. B] {10.1016/j.physletb.2015.11.025}, 752,
  182

\bibitem[\protect\citeauthoryear{Di~Valentino, Melchiorri  \&
  Silk}{Di~Valentino et~al.}{2016b}]{DiValentino:2016hlg}
Di~Valentino E.,  Melchiorri A.,   Silk J.,  2016b, \mn@doi [Phys. Lett. B]
  {10.1016/j.physletb.2016.08.043}, 761, 242

\bibitem[\protect\citeauthoryear{Di~Valentino, Melchiorri  \&
  Silk}{Di~Valentino et~al.}{2019}]{DiValentino:2019qzk}
Di~Valentino E.,  Melchiorri A.,   Silk J.,  2019, \mn@doi [Nature Astron.]
  {10.1038/s41550-019-0906-9}, 4, 196

\bibitem[\protect\citeauthoryear{Di~Valentino, Melchiorri  \&
  Silk}{Di~Valentino et~al.}{2020a}]{DiValentino:2019dzu}
Di~Valentino E.,  Melchiorri A.,   Silk J.,  2020a, \mn@doi [JCAP]
  {10.1088/1475-7516/2020/01/013}, 01, 013

\bibitem[\protect\citeauthoryear{Di~Valentino, Linder  \&
  Melchiorri}{Di~Valentino et~al.}{2020b}]{DiValentino:2020kha}
Di~Valentino E.,  Linder E.~V.,   Melchiorri A.,  2020b, \mn@doi [Phys. Dark
  Univ.] {10.1016/j.dark.2020.100733}, 30, 100733

\bibitem[\protect\citeauthoryear{Di~Valentino et~al.,}{Di~Valentino
  et~al.}{2021a}]{DiValentino:2021izs}
Di~Valentino E.,  et~al., 2021a, {In the Realm of the Hubble tension $-$ a
  Review of Solutions} (\mn@eprint {arXiv} {2103.01183})

\bibitem[\protect\citeauthoryear{Di~Valentino, Gariazzo  \& Mena}{Di~Valentino
  et~al.}{2021b}]{DiValentino:2021hoh}
Di~Valentino E.,  Gariazzo S.,   Mena O.,  2021b, \mn@doi [Phys. Rev. D]
  {10.1103/PhysRevD.104.083504}, 104, 083504

\bibitem[\protect\citeauthoryear{Di~Valentino et~al.}{Di~Valentino
  et~al.}{2021c}]{DiValentino:2020srs}
Di~Valentino E.,  et~al., 2021c, \mn@doi [Astropart. Phys.]
  {10.1016/j.astropartphys.2021.102607}, 131, 102607

\bibitem[\protect\citeauthoryear{Di~Valentino, Melchiorri  \&
  Silk}{Di~Valentino et~al.}{2021d}]{DiValentino:2020hov}
Di~Valentino E.,  Melchiorri A.,   Silk J.,  2021d, \mn@doi [Astrophys. J.
  Lett.] {10.3847/2041-8213/abe1c4}, 908, L9

\bibitem[\protect\citeauthoryear{Dialektopoulos, Said, Mifsud, Sultana  \&
  Adami}{Dialektopoulos et~al.}{2021}]{Dialektopoulos:2021wde}
Dialektopoulos K.,  Said J.~L.,  Mifsud J.,  Sultana J.,   Adami K.~Z.,  2021,
  {Neural Network Reconstruction of Late-Time Cosmology and Null Tests}
  (\mn@eprint {arXiv} {2111.11462})

\bibitem[\protect\citeauthoryear{Efstathiou}{Efstathiou}{2021}]{Efstathiou:2021ocp}
Efstathiou G.,  2021, \mn@doi [Mon. Not. Roy. Astron. Soc.]
  {10.1093/mnras/stab1588}, 505, 3866

\bibitem[\protect\citeauthoryear{Efstathiou \& Gratton}{Efstathiou \&
  Gratton}{2020}]{Efstathiou:2020wem}
Efstathiou G.,  Gratton S.,  2020, \mn@doi [Mon. Not. Roy. Astron. Soc.]
  {10.1093/mnrasl/slaa093}, 496, L91

\bibitem[\protect\citeauthoryear{Ellis, Stoeger, McEwan  \& Dunsby}{Ellis
  et~al.}{2002}]{Ellis:2001ym}
Ellis G. F.~R.,  Stoeger W.~R.,  McEwan P.,   Dunsby P.,  2002, \mn@doi [Gen.
  Rel. Grav.] {10.1023/A:1020087004012}, 34, 1445

\bibitem[\protect\citeauthoryear{Etherington}{Etherington}{1933}]{DDR_original}
Etherington I.,  1933, \mn@doi [The London, Edinburgh, and Dublin Philosophical
  Magazine and Journal of Science] {10.1080/14786443309462220}, 15, 761

\bibitem[\protect\citeauthoryear{Fixsen et~al.}{Fixsen
  et~al.}{1994}]{Fixsen:1993rd}
Fixsen D.~J.,  et~al., 1994, \mn@doi [Astrophys. J.] {10.1086/173575}, 420, 445

\bibitem[\protect\citeauthoryear{Font-Ribera et~al.}{Font-Ribera
  et~al.}{2014}]{BOSS:2013igd}
Font-Ribera A.,  et~al., 2014, \mn@doi [JCAP] {10.1088/1475-7516/2014/05/027},
  05, 027

\bibitem[\protect\citeauthoryear{Forconi, Giar\`e, Di~Valentino  \&
  Melchiorri}{Forconi et~al.}{2021}]{Forconi:2021que}
Forconi M.,  Giar\`e W.,  Di~Valentino E.,   Melchiorri A.,  2021,
  {Cosmological constraints on slow-roll inflation: an update} (\mn@eprint
  {arXiv} {2110.01695})

\bibitem[\protect\citeauthoryear{Freedman}{Freedman}{2021}]{Freedman:2021ahq}
Freedman W.~L.,  2021, \mn@doi [Astrophys. J.] {10.3847/1538-4357/ac0e95}, 919,
  16

\bibitem[\protect\citeauthoryear{Freedman et~al.}{Freedman
  et~al.}{2019}]{Freedman:2019jwv}
Freedman W.~L.,  et~al., 2019, \mn@doi [The Astrophysical Journal]
  {10.3847/1538-4357/ab2f73}

\bibitem[\protect\citeauthoryear{Freese \& Winkler}{Freese \&
  Winkler}{2021}]{Freese:2021rjq}
Freese K.,  Winkler M.~W.,  2021, {Chain Early Dark Energy: Solving the Hubble
  Tension and Explaining Today's Dark Energy} (\mn@eprint {arXiv} {2102.13655})

\bibitem[\protect\citeauthoryear{Froustey, Pitrou  \& Volpe}{Froustey
  et~al.}{2020}]{Froustey:2020mcq}
Froustey J.,  Pitrou C.,   Volpe M.~C.,  2020, \mn@doi [JCAP]
  {10.1088/1475-7516/2020/12/015}, 12, 015

\bibitem[\protect\citeauthoryear{Giar\`e, Renzi, Melchiorri, Mena  \&
  Di~Valentino}{Giar\`e et~al.}{2021a}]{Giare:2021cqr}
Giar\`e W.,  Renzi F.,  Melchiorri A.,  Mena O.,   Di~Valentino E.,  2021a,
  {Cosmological forecasts on thermal axions, relic neutrinos and light
  elements} (\mn@eprint {arXiv} {2110.00340})

\bibitem[\protect\citeauthoryear{Giar\`e, Di~Valentino, Melchiorri  \&
  Mena}{Giar\`e et~al.}{2021b}]{Giare:2020vzo}
Giar\`e W.,  Di~Valentino E.,  Melchiorri A.,   Mena O.,  2021b, \mn@doi [Mon.
  Not. Roy. Astron. Soc.] {10.1093/mnras/stab1442}, 505, 2703

\bibitem[\protect\citeauthoryear{Gil-Marin et~al.}{Gil-Marin
  et~al.}{2020}]{Gil-Marin:2020bct}
Gil-Marin H.,  et~al., 2020, \mn@doi [MNRAS] {10.1093/mnras/staa2455}, 498,
  2492

\bibitem[\protect\citeauthoryear{Giusarma, Gerbino, Mena, Vagnozzi, Ho  \&
  Freese}{Giusarma et~al.}{2016}]{Giusarma:2016phn}
Giusarma E.,  Gerbino M.,  Mena O.,  Vagnozzi S.,  Ho S.,   Freese K.,  2016,
  \mn@doi [Phys. Rev. D] {10.1103/PhysRevD.94.083522}, 94, 083522

\bibitem[\protect\citeauthoryear{Gordon, Li  \& Singh}{Gordon
  et~al.}{2021}]{Gordon:2020gel}
Gordon L.,  Li B.-F.,   Singh P.,  2021, \mn@doi [Phys. Rev. D]
  {10.1103/PhysRevD.103.046016}, 103, 046016

\bibitem[\protect\citeauthoryear{Green \& Meyers}{Green \&
  Meyers}{2021}]{Green:2021xzn}
Green D.,  Meyers J.,  2021, {Cosmological Implications of a Neutrino Mass
  Detection} (\mn@eprint {arXiv} {2111.01096})

\bibitem[\protect\citeauthoryear{Green, Guo  \& Wallisch}{Green
  et~al.}{2021}]{Green:2021hjh}
Green D.,  Guo Y.,   Wallisch B.,  2021, {Cosmological Implications of
  Axion-Matter Couplings} (\mn@eprint {arXiv} {2109.12088})

\bibitem[\protect\citeauthoryear{Hagstotz, de Salas, Gariazzo, Gerbino,
  Lattanzi, Vagnozzi, Freese  \& Pastor}{Hagstotz
  et~al.}{2020}]{Hagstotz:2020ukm}
Hagstotz S.,  de Salas P.~F.,  Gariazzo S.,  Gerbino M.,  Lattanzi M.,
  Vagnozzi S.,  Freese K.,   Pastor S.,  2020, {Bounds on light sterile
  neutrino mass and mixing from cosmology and laboratory searches} (\mn@eprint
  {arXiv} {2003.02289})

\bibitem[\protect\citeauthoryear{Handley}{Handley}{2019}]{Handley:2019anl}
Handley W.,  2019, \mn@doi [Phys. Rev. D] {10.1103/PhysRevD.100.123517}, 100,
  123517

\bibitem[\protect\citeauthoryear{Handley}{Handley}{2021}]{Handley:2019tkm}
Handley W.,  2021, \mn@doi [Phys. Rev. D] {10.1103/PhysRevD.103.L041301}, 103,
  L041301

\bibitem[\protect\citeauthoryear{Hogg, Martinelli  \& Nesseris}{Hogg
  et~al.}{2020}]{Hogg:2020ktc}
Hogg N.~B.,  Martinelli M.,   Nesseris S.,  2020, \mn@doi [JCAP]
  {10.1088/1475-7516/2020/12/019}, 12, 019

\bibitem[\protect\citeauthoryear{Holanda \& da Silva}{Holanda \&
  da~Silva}{2020}]{Holanda:2020fmo}
Holanda R.,  da Silva W.,  2020, {On a possible cosmological evolution of
  galaxy cluster $Y_{\rm X}-Y_{\rm SZE}$ scaling relation} (\mn@eprint {arXiv}
  {2007.14199})

\bibitem[\protect\citeauthoryear{Holanda, Busti  \& Alcaniz}{Holanda
  et~al.}{2016}]{Holanda:2015zpz}
Holanda R.,  Busti V.,   Alcaniz J.,  2016, \mn@doi [JCAP]
  {10.1088/1475-7516/2016/02/054}, 02, 054

\bibitem[\protect\citeauthoryear{Holanda, Busti, Lima  \& Alcaniz}{Holanda
  et~al.}{2017}]{Holanda:2016msr}
Holanda R.,  Busti V.,  Lima F.,   Alcaniz J.,  2017, \mn@doi [JCAP]
  {10.1088/1475-7516/2017/09/039}, 09, 039

\bibitem[\protect\citeauthoryear{Hou et~al.}{Hou et~al.}{2020}]{Hou:2020rse}
Hou J.,  et~al., 2020, \mn@doi [MNRAS] {10.1093/mnras/staa3234}, 500, 1201

\bibitem[\protect\citeauthoryear{Huang et~al.}{Huang
  et~al.}{2018}]{Huang:2018dbn}
Huang C.~D.,  et~al., 2018, \mn@doi [The Astrophysical Journal]
  {10.3847/1538-4357/aab6b3}, 857, 67

\bibitem[\protect\citeauthoryear{Huang et~al.}{Huang
  et~al.}{2020}]{Huang:2019yhh}
Huang C.~D.,  et~al., 2020, \mn@doi [The Astrophysical Journal]
  {10.3847/1538-4357/ab5dbd}, 889, 5

\bibitem[\protect\citeauthoryear{Ivanov, Ali-Ha\"\i{}moud  \&
  Lesgourgues}{Ivanov et~al.}{2020}]{Ivanov:2020mfr}
Ivanov M.~M.,  Ali-Ha\"\i{}moud Y.,   Lesgourgues J.,  2020, \mn@doi [Phys.
  Rev. D] {10.1103/PhysRevD.102.063515}, 102, 063515

\bibitem[\protect\citeauthoryear{Kazantzidis, Koo, Nesseris, Perivolaropoulos
  \& Shafieloo}{Kazantzidis et~al.}{2021}]{Kazantzidis:2020xta}
Kazantzidis L.,  Koo H.,  Nesseris S.,  Perivolaropoulos L.,   Shafieloo A.,
  2021, \mn@doi [Mon. Not. Roy. Astron. Soc.] {10.1093/mnras/staa3866}, 501,
  3421

\bibitem[\protect\citeauthoryear{Knox \& Millea}{Knox \&
  Millea}{2020}]{Knox:2019rjx}
Knox L.,  Millea M.,  2020, \mn@doi [Phys. Rev. D]
  {10.1103/PhysRevD.101.043533}, 101, 043533

\bibitem[\protect\citeauthoryear{Krishnan, Mohayaee, \'O~Colg\'ain,
  Sheikh-Jabbari  \& Yin}{Krishnan et~al.}{2021a}]{Krishnan:2021jmh}
Krishnan C.,  Mohayaee R.,  \'O~Colg\'ain E.,  Sheikh-Jabbari M.~M.,   Yin L.,
  2021a, {Hints of FLRW Breakdown from Supernovae} (\mn@eprint {arXiv}
  {2106.02532})

\bibitem[\protect\citeauthoryear{Krishnan, Mohayaee, Colg\'ain, Sheikh-Jabbari
  \& Yin}{Krishnan et~al.}{2021b}]{Krishnan:2021dyb}
Krishnan C.,  Mohayaee R.,  Colg\'ain E.~O.,  Sheikh-Jabbari M.~M.,   Yin L.,
  2021b, \mn@doi [Class. Quant. Grav.] {10.1088/1361-6382/ac1a81}, 38, 184001

\bibitem[\protect\citeauthoryear{Lewis}{Lewis}{2013}]{Lewis:2013hha}
Lewis A.,  2013, \mn@doi [Phys. Rev.] {10.1103/PhysRevD.87.103529}, D87, 103529

\bibitem[\protect\citeauthoryear{Lewis \& Bridle}{Lewis \&
  Bridle}{2002}]{Lewis:2002ah}
Lewis A.,  Bridle S.,  2002, \mn@doi [Phys. Rev.] {10.1103/PhysRevD.66.103511},
  D66, 103511

\bibitem[\protect\citeauthoryear{Li, Wu  \& Yu}{Li et~al.}{2011}]{Li:2011exa}
Li Z.,  Wu P.,   Yu H.~W.,  2011, \mn@doi [Astrophys. J. Lett.]
  {10.1088/2041-8205/729/1/L14}, 729, L14

\bibitem[\protect\citeauthoryear{Liao}{Liao}{2019}]{Liao:2019xug}
Liao K.,  2019, \mn@doi [Astrophys. J.] {10.3847/1538-4357/ab4819}, 885, 70

\bibitem[\protect\citeauthoryear{Lin, Li  \& Tang}{Lin
  et~al.}{2021}]{Lin:2020vqj}
Lin H.-N.,  Li X.,   Tang L.,  2021, \mn@doi [Chin. Phys. C]
  {10.1088/1674-1137/abc53a}, 45, 015109

\bibitem[\protect\citeauthoryear{Linde}{Linde}{1995}]{Linde:1995xm}
Linde A.~D.,  1995, \mn@doi [Phys. Lett. B] {10.1016/0370-2693(95)00370-Z},
  351, 99

\bibitem[\protect\citeauthoryear{Linde}{Linde}{2003}]{Linde:2003hc}
Linde A.~D.,  2003, \mn@doi [JCAP] {10.1088/1475-7516/2003/05/002}, 05, 002

\bibitem[\protect\citeauthoryear{Luongo, Muccino, Colg\'ain, Sheikh-Jabbari  \&
  Yin}{Luongo et~al.}{2021}]{Luongo:2021nqh}
Luongo O.,  Muccino M.,  Colg\'ain E.~O.,  Sheikh-Jabbari M.~M.,   Yin L.,
  2021, {On Larger $H_0$ Values in the CMB Dipole Direction} (\mn@eprint
  {arXiv} {2108.13228})

\bibitem[\protect\citeauthoryear{Ma \& Corasaniti}{Ma \&
  Corasaniti}{2018}]{Ma:2016bjt}
Ma C.,  Corasaniti P.-S.,  2018, \mn@doi [Astrophys. J.]
  {10.3847/1538-4357/aac88f}, 861, 124

\bibitem[\protect\citeauthoryear{Mangano, Miele, Pastor, Pinto, Pisanti  \&
  Serpico}{Mangano et~al.}{2005}]{Mangano:2005cc}
Mangano G.,  Miele G.,  Pastor S.,  Pinto T.,  Pisanti O.,   Serpico P.~D.,
  2005, \mn@doi [Nucl. Phys. B] {10.1016/j.nuclphysb.2005.09.041}, 729, 221

\bibitem[\protect\citeauthoryear{Marra \& Perivolaropoulos}{Marra \&
  Perivolaropoulos}{2021}]{Marra:2021fvf}
Marra V.,  Perivolaropoulos L.,  2021, \mn@doi [Phys. Rev. D]
  {10.1103/PhysRevD.104.L021303}, 104, L021303

\bibitem[\protect\citeauthoryear{Martinelli \& Tutusaus}{Martinelli \&
  Tutusaus}{2019}]{Martinelli:2019krf}
Martinelli M.,  Tutusaus I.,  2019, \mn@doi [Symmetry] {10.3390/sym11080986},
  11, 986

\bibitem[\protect\citeauthoryear{Martinelli, Hogg, Peirone, Bruni  \&
  Wands}{Martinelli et~al.}{2019}]{Martinelli:2019dau}
Martinelli M.,  Hogg N.~B.,  Peirone S.,  Bruni M.,   Wands D.,  2019, \mn@doi
  [Mon. Not. Roy. Astron. Soc.] {10.1093/mnras/stz1915}, 488, 3423

\bibitem[\protect\citeauthoryear{Martinelli et~al.}{Martinelli
  et~al.}{2020}]{EUCLID:2020syl}
Martinelli M.,  et~al., 2020, \mn@doi [Astron. Astrophys.]
  {10.1051/0004-6361/202039078}, 644, A80

\bibitem[\protect\citeauthoryear{Masaki, Aoki  \& Soda}{Masaki
  et~al.}{2017}]{Masaki:2017aea}
Masaki E.,  Aoki A.,   Soda J.,  2017, \mn@doi [Phys. Rev. D]
  {10.1103/PhysRevD.96.043519}, 96, 043519

\bibitem[\protect\citeauthoryear{Masaki, Aoki  \& Soda}{Masaki
  et~al.}{2020}]{Masaki:2019ggg}
Masaki E.,  Aoki A.,   Soda J.,  2020, \mn@doi [Phys. Rev. D]
  {10.1103/PhysRevD.101.043505}, 101, 043505

\bibitem[\protect\citeauthoryear{McClure \& Dyer}{McClure \&
  Dyer}{2007}]{McClure:2007vv}
McClure M.~L.,  Dyer C.~C.,  2007, \mn@doi [New Astron.]
  {10.1016/j.newast.2007.03.005}, 12, 533

\bibitem[\protect\citeauthoryear{Melchiorri, Mena  \& Slosar}{Melchiorri
  et~al.}{2007}]{Melchiorri:2007cd}
Melchiorri A.,  Mena O.,   Slosar A.,  2007, \mn@doi [Phys. Rev. D]
  {10.1103/PhysRevD.76.041303}, 76, 041303

\bibitem[\protect\citeauthoryear{Mirizzi, Raffelt  \& Serpico}{Mirizzi
  et~al.}{2008}]{Mirizzi:2006zy}
Mirizzi A.,  Raffelt G.~G.,   Serpico P.~D.,  2008, \mn@doi [Lect. Notes Phys.]
  {10.1007/978-3-540-73518-2_7}, 741, 115

\bibitem[\protect\citeauthoryear{Moresco}{Moresco}{2015}]{Moresco:2015cya}
Moresco M.,  2015, \mn@doi [MNRAS] {10.1093/mnrasl/slv037}, 450, L16

\bibitem[\protect\citeauthoryear{Moresco, Verde, Pozzetti, Jimenez  \&
  Cimatti}{Moresco et~al.}{2012a}]{Moresco:2012by}
Moresco M.,  Verde L.,  Pozzetti L.,  Jimenez R.,   Cimatti A.,  2012a, \mn@doi
  [JCAP] {10.1088/1475-7516/2012/07/053}, 07, 053

\bibitem[\protect\citeauthoryear{{Moresco} et~al.}{{Moresco}
  et~al.}{2012b}]{Moresco:2012}
{Moresco} M.,  et~al., 2012b, \mn@doi [JCAP] {10.1088/1475-7516/2012/08/006},
  \href {https://ui.adsabs.harvard.edu/abs/2012JCAP...08..006M} {2012, 006}

\bibitem[\protect\citeauthoryear{Moresco et~al.,}{Moresco
  et~al.}{2016}]{Moresco:2016mzx}
Moresco M.,  et~al., 2016, \mn@doi [JCAP] {10.1088/1475-7516/2016/05/014}, 05,
  014

\bibitem[\protect\citeauthoryear{Moresco, Jimenez, Verde, Pozzetti, Cimatti  \&
  Citro}{Moresco et~al.}{2018}]{Moresco:2018xdr}
Moresco M.,  Jimenez R.,  Verde L.,  Pozzetti L.,  Cimatti A.,   Citro A.,
  2018, \mn@doi [Astrophys. J.] {10.3847/1538-4357/aae829}, 868, 84

\bibitem[\protect\citeauthoryear{Moresco, Jimenez, Verde, Cimatti  \&
  Pozzetti}{Moresco et~al.}{2020}]{Moresco:2020fbm}
Moresco M.,  Jimenez R.,  Verde L.,  Cimatti A.,   Pozzetti L.,  2020, \mn@doi
  [Astrophys. J.] {10.3847/1538-4357/ab9eb0}, 898, 82

\bibitem[\protect\citeauthoryear{Moresco et~al.}{Moresco
  et~al.}{2022}]{Moresco:2022phi}
Moresco M.,  et~al., 2022, {Unveiling the Universe with Emerging Cosmological
  Probes} (\mn@eprint {arXiv} {2201.07241})

\bibitem[\protect\citeauthoryear{Mortsell, Goobar, Johansson  \&
  Dhawan}{Mortsell et~al.}{2021}]{Mortsell:2021nzg}
Mortsell E.,  Goobar A.,  Johansson J.,   Dhawan S.,  2021, {The Hubble Tension
  Bites the Dust: Sensitivity of the Hubble Constant Determination to Cepheid
  Color Calibration} (\mn@eprint {arXiv} {2105.11461})

\bibitem[\protect\citeauthoryear{Motaharfar \& Singh}{Motaharfar \&
  Singh}{2021}]{Motaharfar:2021gwi}
Motaharfar M.,  Singh P.,  2021, {On the role of dissipative effects in the
  quantum gravitational onset of warm Starobinsky inflation in a closed
  universe} (\mn@eprint {arXiv} {2102.09578})

\bibitem[\protect\citeauthoryear{Mukherjee, Khatri  \& Wandelt}{Mukherjee
  et~al.}{2019}]{Mukherjee:2018zzg}
Mukherjee S.,  Khatri R.,   Wandelt B.~D.,  2019, \mn@doi [JCAP]
  {10.1088/1475-7516/2019/06/031}, 06, 031

\bibitem[\protect\citeauthoryear{{Neal}}{{Neal}}{2005}]{Neal:2005}
{Neal} R.~M.,  2005, arXiv Mathematics e-prints, \href
  {https://ui.adsabs.harvard.edu/abs/2005math......2099N} {p. math/0502099}

\bibitem[\protect\citeauthoryear{Neveux et~al.}{Neveux
  et~al.}{2020}]{Neveux:2020voa}
Neveux R.,  et~al., 2020, \mn@doi [MNRAS] {10.1093/mnras/staa2780}, 499, 210

\bibitem[\protect\citeauthoryear{Niedermann \& Sloth}{Niedermann \&
  Sloth}{2020}]{Niedermann:2020dwg}
Niedermann F.,  Sloth M.~S.,  2020, \mn@doi [Phys. Rev. D]
  {10.1103/PhysRevD.102.063527}, 102, 063527

\bibitem[\protect\citeauthoryear{Ooba, Ratra  \& Sugiyama}{Ooba
  et~al.}{2018}]{Ooba:2017ukj}
Ooba J.,  Ratra B.,   Sugiyama N.,  2018, \mn@doi [Astrophys. J.]
  {10.3847/1538-4357/aad633}, 864, 80

\bibitem[\protect\citeauthoryear{Peirone, Benevento, Frusciante  \&
  Tsujikawa}{Peirone et~al.}{2019}]{Peirone:2019aua}
Peirone S.,  Benevento G.,  Frusciante N.,   Tsujikawa S.,  2019, \mn@doi
  [Phys. Rev. D] {10.1103/PhysRevD.100.063540}, 100, 063540

\bibitem[\protect\citeauthoryear{Perivolaropoulos \& Skara}{Perivolaropoulos \&
  Skara}{2021}]{Perivolaropoulos:2021jda}
Perivolaropoulos L.,  Skara F.,  2021, {Challenges for $\Lambda$CDM: An update}
  (\mn@eprint {arXiv} {2105.05208})

\bibitem[\protect\citeauthoryear{Pogosian, Raveri, Koyama, Martinelli,
  Silvestri  \& Zhao}{Pogosian et~al.}{2021}]{Pogosian:2021mcs}
Pogosian L.,  Raveri M.,  Koyama K.,  Martinelli M.,  Silvestri A.,   Zhao
  G.-B.,  2021, {Imprints of cosmological tensions in reconstructed gravity}
  (\mn@eprint {arXiv} {2107.12992})

\bibitem[\protect\citeauthoryear{Poulin, Smith, Karwal  \& Kamionkowski}{Poulin
  et~al.}{2019}]{Poulin:2018cxd}
Poulin V.,  Smith T.~L.,  Karwal T.,   Kamionkowski M.,  2019, \mn@doi
  [Physical Review Letters] {10.1103/PhysRevLett.122.221301}, 122, 221301

\bibitem[\protect\citeauthoryear{Powell}{Powell}{2009}]{BOBYQA:3}
Powell M.,  2009, {Escaping local minima with derivative-free methods: a
  numerical investigation}.
Tech. Rep. DAMTP 2009/NA06, University of Cambridge,

\bibitem[\protect\citeauthoryear{Rana, Jain, Mahajan, Mukherjee  \&
  Holanda}{Rana et~al.}{2017}]{Rana:2017sfr}
Rana A.,  Jain D.,  Mahajan S.,  Mukherjee A.,   Holanda R.,  2017, \mn@doi
  [JCAP] {10.1088/1475-7516/2017/07/010}, 07, 010

\bibitem[\protect\citeauthoryear{Rasmussen \& Williams}{Rasmussen \&
  Williams}{2006}]{GPbible}
Rasmussen C.~E.,  Williams C. K.~I.,  2006, Gaussian Processes for Machine
  Learning.
MIT Press

\bibitem[\protect\citeauthoryear{Ratra}{Ratra}{2017}]{Ratra:2017ezv}
Ratra B.,  2017, \mn@doi [Phys. Rev. D] {10.1103/PhysRevD.96.103534}, 96,
  103534

\bibitem[\protect\citeauthoryear{Ratsimbazafy, Loubser, Crawford, Cress,
  Bassett, Nichol  \& V\"ais\"anen}{Ratsimbazafy
  et~al.}{2017}]{Ratsimbazafy:2017vga}
Ratsimbazafy A.,  Loubser S.,  Crawford S.,  Cress C.,  Bassett B.,  Nichol R.,
    V\"ais\"anen P.,  2017, \mn@doi [MNRAS] {10.1093/mnras/stx301}, 467, 3239

\bibitem[\protect\citeauthoryear{Raveri}{Raveri}{2020}]{Raveri:2019mxg}
Raveri M.,  2020, \mn@doi [Phys. Rev. D] {10.1103/PhysRevD.101.083524}, 101,
  083524

\bibitem[\protect\citeauthoryear{Raveri et~al.,}{Raveri
  et~al.}{2021}]{Raveri:2021dbu}
Raveri M.,  et~al., 2021, {A joint reconstruction of dark energy and modified
  growth evolution} (\mn@eprint {arXiv} {2107.12990})

\bibitem[\protect\citeauthoryear{Renzi \& Silvestri}{Renzi \&
  Silvestri}{2020}]{Renzi:2020fnx}
Renzi F.,  Silvestri A.,  2020, {A look at the Hubble speed from first
  principles} (\mn@eprint {arXiv} {2011.10559})

\bibitem[\protect\citeauthoryear{Renzi, Hogg, Martinelli  \& Nesseris}{Renzi
  et~al.}{2021}]{Renzi:2020bvl}
Renzi F.,  Hogg N.~B.,  Martinelli M.,   Nesseris S.,  2021, \mn@doi [Phys.
  Dark Univ.] {10.1016/j.dark.2021.100824}, 32, 100824

\bibitem[\protect\citeauthoryear{Riess et~al.}{Riess
  et~al.}{2016}]{Riess:2016jrr}
Riess A.~G.,  et~al., 2016, \mn@doi [Astrophys. J.]
  {10.3847/0004-637X/826/1/56}, 826, 56

\bibitem[\protect\citeauthoryear{Riess et~al.}{Riess
  et~al.}{2018}]{Riess:2017lxs}
Riess A.~G.,  et~al., 2018, \mn@doi [Astrophys. J.] {10.3847/1538-4357/aaa5a9},
  853, 126

\bibitem[\protect\citeauthoryear{Riess et~al.}{Riess
  et~al.}{2021a}]{Riess:2021jrx}
Riess A.~G.,  et~al., 2021a, {A Comprehensive Measurement of the Local Value of
  the Hubble Constant with 1 km/s/Mpc Uncertainty from the Hubble Space
  Telescope and the SH0ES Team} (\mn@eprint {arXiv} {2112.04510})

\bibitem[\protect\citeauthoryear{Riess, Casertano, Yuan, Bowers, Macri, Zinn
  \& Scolnic}{Riess et~al.}{2021b}]{Riess:2020fzl}
Riess A.~G.,  Casertano S.,  Yuan W.,  Bowers J.~B.,  Macri L.,  Zinn J.~C.,
  Scolnic D.,  2021b, \mn@doi [Astrophys. J. Lett.] {10.3847/2041-8213/abdbaf},
  908, L6

\bibitem[\protect\citeauthoryear{Salvatelli, Said, Bruni, Melchiorri  \&
  Wands}{Salvatelli et~al.}{2014}]{Salvatelli2014}
Salvatelli V.,  Said N.,  Bruni M.,  Melchiorri A.,   Wands D.,  2014, \mn@doi
  [Physical Review Letters] {10.1103/PhysRevLett.113.181301}, 113, 181301

\bibitem[\protect\citeauthoryear{Schaefer}{Schaefer}{1999}]{Schaefer:1998zg}
Schaefer B.~E.,  1999, \mn@doi [Phys. Rev. Lett.]
  {10.1103/PhysRevLett.82.4964}, 82, 4964

\bibitem[\protect\citeauthoryear{Schwarz}{Schwarz}{2003}]{Schwarz:2003du}
Schwarz D.~J.,  2003, \mn@doi [Annalen Phys.] {10.1002/andp.200310010}, 12, 220

\bibitem[\protect\citeauthoryear{Scolnic et~al.}{Scolnic
  et~al.}{2018}]{Scolnic:2017caz}
Scolnic D.~M.,  et~al., 2018, \mn@doi [Astrophys. J.]
  {10.3847/1538-4357/aab9bb}, 859, 101

\bibitem[\protect\citeauthoryear{Scolnic et~al.}{Scolnic
  et~al.}{2021}]{Scolnic:2021amr}
Scolnic D.,  et~al., 2021, {The Pantheon+ Type Ia Supernova Sample: The Full
  Dataset and Light-Curve Release} (\mn@eprint {arXiv} {2112.03863})

\bibitem[\protect\citeauthoryear{Sloan, Dimopoulos  \& Karamitsos}{Sloan
  et~al.}{2020}]{Sloan:2019jyl}
Sloan D.,  Dimopoulos K.,   Karamitsos S.,  2020, \mn@doi [Phys. Rev. D]
  {10.1103/PhysRevD.101.043521}, 101, 043521

\bibitem[\protect\citeauthoryear{{Stern}, {Jimenez}, {Verde}, {Kamionkowski}
  \& {Stanford}}{{Stern} et~al.}{2010}]{Stern:2010}
{Stern} D.,  {Jimenez} R.,  {Verde} L.,  {Kamionkowski} M.,   {Stanford} S.~A.,
   2010, \mn@doi [JCAP] {10.1088/1475-7516/2010/02/008}, \href
  {https://ui.adsabs.harvard.edu/abs/2010JCAP...02..008S} {2010, 008}

\bibitem[\protect\citeauthoryear{Tamone et~al.}{Tamone
  et~al.}{2020}]{Tamone:2020qrl}
Tamone A.,  et~al., 2020, \mn@doi [MNRAS] {10.1093/mnras/staa3050}, 499, 5527

\bibitem[\protect\citeauthoryear{Tiwari}{Tiwari}{2017}]{Tiwari:2016cps}
Tiwari P.,  2017, \mn@doi [Phys. Rev. D] {10.1103/PhysRevD.95.023005}, 95,
  023005

\bibitem[\protect\citeauthoryear{Torrado \& Lewis}{Torrado \&
  Lewis}{2021}]{Torrado:2020dgo}
Torrado J.,  Lewis A.,  2021, \mn@doi [JCAP] {10.1088/1475-7516/2021/05/057},
  05, 057

\bibitem[\protect\citeauthoryear{Unger \& Pop\l{}awski}{Unger \&
  Pop\l{}awski}{2019}]{Unger:2018oqo}
Unger G.,  Pop\l{}awski N.,  2019, \mn@doi [Astrophys. J.]
  {10.3847/1538-4357/aaf169}, 870, 78

\bibitem[\protect\citeauthoryear{Uzan, Kirchner  \& Ellis}{Uzan
  et~al.}{2003}]{Uzan:2003nk}
Uzan J.-P.,  Kirchner U.,   Ellis G. F.~R.,  2003, \mn@doi [Mon. Not. Roy.
  Astron. Soc.] {10.1046/j.1365-8711.2003.07043.x}, 344, L65

\bibitem[\protect\citeauthoryear{Vagnozzi}{Vagnozzi}{2019}]{Vagnozzi:2019utt}
Vagnozzi S.,  2019, arxiv:1907.08010 [astro-ph.CO]

\bibitem[\protect\citeauthoryear{Vagnozzi, Giusarma, Mena, Freese, Gerbino, Ho
  \& Lattanzi}{Vagnozzi et~al.}{2017}]{Vagnozzi:2017ovm}
Vagnozzi S.,  Giusarma E.,  Mena O.,  Freese K.,  Gerbino M.,  Ho S.,
  Lattanzi M.,  2017, \mn@doi [Phys. Rev. D] {10.1103/PhysRevD.96.123503}, 96,
  123503

\bibitem[\protect\citeauthoryear{Vagnozzi, Brinckmann, Archidiacono, Freese,
  Gerbino, Lesgourgues  \& Sprenger}{Vagnozzi et~al.}{2018a}]{Vagnozzi:2018pwo}
Vagnozzi S.,  Brinckmann T.,  Archidiacono M.,  Freese K.,  Gerbino M.,
  Lesgourgues J.,   Sprenger T.,  2018a, \mn@doi [JCAP]
  {10.1088/1475-7516/2018/09/001}, 09, 001

\bibitem[\protect\citeauthoryear{Vagnozzi, Dhawan, Gerbino, Freese, Goobar  \&
  Mena}{Vagnozzi et~al.}{2018b}]{Vagnozzi:2018jhn}
Vagnozzi S.,  Dhawan S.,  Gerbino M.,  Freese K.,  Goobar A.,   Mena O.,
  2018b, \mn@doi [Phys. Rev. D] {10.1103/PhysRevD.98.083501}, 98, 083501

\bibitem[\protect\citeauthoryear{Vagnozzi, Di~Valentino, Gariazzo, Melchiorri,
  Mena  \& Silk}{Vagnozzi et~al.}{2020}]{Vagnozzi:2020zrh}
Vagnozzi S.,  Di~Valentino E.,  Gariazzo S.,  Melchiorri A.,  Mena O.,   Silk
  J.,  2020, {Listening to the BOSS: the galaxy power spectrum take on spatial
  curvature and cosmic concordance} (\mn@eprint {arXiv} {2010.02230})

\bibitem[\protect\citeauthoryear{Vagnozzi, Loeb  \& Moresco}{Vagnozzi
  et~al.}{2021}]{Vagnozzi:2020dfn}
Vagnozzi S.,  Loeb A.,   Moresco M.,  2021, \mn@doi [Astrophys. J.]
  {10.3847/1538-4357/abd4df}, 908, 84

\bibitem[\protect\citeauthoryear{Verde, Treu  \& Riess}{Verde
  et~al.}{2019}]{Verde:2019ivm}
Verde L.,  Treu T.,   Riess A.~G.,  2019, \mn@doi [Nature Astron.]
  {10.1038/s41550-019-0902-0}, 3, 891

\bibitem[\protect\citeauthoryear{{Vogl}}{{Vogl}}{2020}]{TypeIISNa}
{Vogl} C.,  2020, PhD thesis, Technical University of Munich, Germany

\bibitem[\protect\citeauthoryear{Weinberg}{Weinberg}{1972}]{Weinberg}
Weinberg S.,  1972, {Gravitation and Cosmology}.
Wiley, New York, NY, \url {https://cds.cern.ch/record/100595}

\bibitem[\protect\citeauthoryear{Wu et~al.,}{Wu et~al.}{2016}]{Wu:2016brq}
Wu X.-F.,  et~al., 2016, \mn@doi [Astrophys. J. Lett.]
  {10.3847/2041-8205/822/1/L15}, 822, L15

\bibitem[\protect\citeauthoryear{Xu \& Huang}{Xu \& Huang}{2020}]{Xu:2020fxj}
Xu B.,  Huang Q.,  2020, \mn@doi [Eur. Phys. J. Plus]
  {10.1140/epjp/s13360-020-00444-2}, 135, 447

\bibitem[\protect\citeauthoryear{Xu, DePorzio, Mu\~noz  \& Dvorkin}{Xu
  et~al.}{2021}]{Xu:2020fyg}
Xu W.~L.,  DePorzio N.,  Mu\~noz J.~B.,   Dvorkin C.,  2021, \mn@doi [Phys.
  Rev. D] {10.1103/PhysRevD.103.023503}, 103, 023503

\bibitem[\protect\citeauthoryear{{Zhang}, {Zhang}, {Yuan}, {Liu}, {Zhang}  \&
  {Sun}}{{Zhang} et~al.}{2014}]{Zhang:2014}
{Zhang} C.,  {Zhang} H.,  {Yuan} S.,  {Liu} S.,  {Zhang} T.-J.,   {Sun} Y.-C.,
  2014, \mn@doi [Res. Astron. and Astrophys.] {10.1088/1674-4527/14/10/002},
  \href {https://ui.adsabs.harvard.edu/abs/2014RAA....14.1221Z} {14, 1221}

\bibitem[\protect\citeauthoryear{Zhao et~al.}{Zhao et~al.}{2017}]{Zhao:2017cud}
Zhao G.-B.,  et~al., 2017, \mn@doi [Nature Astron.]
  {10.1038/s41550-017-0216-z}, 1, 627

\bibitem[\protect\citeauthoryear{Zhou, Hu, Li, Zhang  \& Fang}{Zhou
  et~al.}{2020}]{Zhou:2020moc}
Zhou C.,  Hu J.,  Li M.,  Zhang X.,   Fang G.,  2020, {A Distance-Deviation
  Consistency and Model-Independent Method to Test the Cosmic Distance-Duality
  Relation} (\mn@eprint {arXiv} {2011.06881})

\bibitem[\protect\citeauthoryear{Zuckerman \& Anchordoqui}{Zuckerman \&
  Anchordoqui}{2021}]{Zuckerman:2021kgm}
Zuckerman E.,  Anchordoqui L.~A.,  2021, {Spatial curvature sensitivity to
  local $H_0$ from the Cepheid distance ladder} (\mn@eprint {arXiv}
  {2110.05346})

\bibitem[\protect\citeauthoryear{de Mattia et~al.}{de~Mattia
  et~al.}{2020}]{deMattia:2020fkb}
de Mattia A.,  et~al., 2020, \mn@doi [MNRAS] {10.1093/mnras/staa3891}, 501,
  5616

\bibitem[\protect\citeauthoryear{de Sainte~Agathe et~al.}{de~Sainte~Agathe
  et~al.}{2019}]{Agathe:2019vsu}
de Sainte~Agathe V.,  et~al., 2019, \mn@doi [Astron. Astrophys.]
  {10.1051/0004-6361/201935638}, 629, A85

\bibitem[\protect\citeauthoryear{de Salas, Forero, Gariazzo,
  Mart\'\i{}nez-Mirav\'e, Mena, Ternes, T\'ortola  \& Valle}{de~Salas
  et~al.}{2021}]{deSalas:2020pgw}
de Salas P.~F.,  Forero D.~V.,  Gariazzo S.,  Mart\'\i{}nez-Mirav\'e P.,  Mena
  O.,  Ternes C.~A.,  T\'ortola M.,   Valle J. W.~F.,  2021, \mn@doi [JHEP]
  {10.1007/JHEP02(2021)071}, 02, 071

\bibitem[\protect\citeauthoryear{du~Mas~des Bourboux et~al.}{du~Mas~des
  Bourboux et~al.}{2017}]{duMasdesBourboux:2017mrl}
du~Mas~des Bourboux H.,  et~al., 2017, \mn@doi [Astron. Astrophys.]
  {10.1051/0004-6361/201731731}, 608, A130

\makeatother
\end{thebibliography}

\appendix
\section{Cosmological Data}\label{sec.AppendixA}
The  baryon acoustic oscillation and cosmic chronometer data used in this work are reported in \autoref{tab:BAOdata} and \autoref{tab:CCdata}. The Pantheon dataset \citep{Scolnic:2017caz} is publicly available at \href{https://github.com/dscolnic/Pantheon}{https://github.com/dscolnic/Pantheon}.

\begin{table*}
\centering
\begin{tabular}{Slcccl}
\toprule
Type & $z$ & $d_M/r_s$ & $d_H/r_s$  & Reference  \\
\hline\hline
BOSS galaxy--galaxy & $ 0.38 $ & $ 10.27\pm 0.15 $ & $ 24.89\pm 0.58 $ & \cite{Alam:2016hwk} \\
eBOSS galaxy--galaxy & $ 0.51 $ & $ 13.38\pm0.18 $ & $ 22.43 \pm 0.48 $ & \cite{Alam:2016hwk}\\
                  & $ 0.70$ & $ 17.65 \pm 0.30 $ & $ 19.78\pm 0.46 $ & \cite{Bautista:2020ahg,Gil-Marin:2020bct}\\
                  & $ 0.85 $ & $ 19.50\pm 1.00 $ & $ 19.60\pm 2.10 $ & \cite{Tamone:2020qrl,deMattia:2020fkb}\\
                  & $1.48$ & $ 30.21\pm0.79 $ & $ 13.23\pm 0.47 $ &\cite{Neveux:2020voa,Hou:2020rse}\\
eBOSS Ly-$\alpha$--Ly-$\alpha$ & $ 2.34 $ & $ 37.41 \pm 1.86 $ & $ 8.86\pm0.29 $ &\cite{Agathe:2019vsu} \\
eBOSS Ly-$\alpha$--quasar & $ 2.35 $ & $ 36.30\pm1.80 $ & $ 8.20\pm0.36 $ & \cite{Blomqvist:2019rah} \\
\bottomrule
\end{tabular}
\caption{The BAO data used in this work given in terms of $d_M = (1+z)d_A$ and $ d_H = c\,H(z)^{-1}$.}
\label{tab:BAOdata}
\end{table*}

\begin{table*}
\centering
\begin{tabular}{Slcllcl}
\toprule
$z$ &  $H(z)$ & Reference & $z$ &  $H(z)$ & Reference  \\
\hline\hline
$0.07$ &  $69.0 \pm 19.6$ & \cite{Zhang:2014} & $0.4783$ &  $80.9 \pm 9.0$ & \cite{Moresco:2016mzx} \\ 
$0.09$ &  $69.0 \pm 12.0$ & \cite{Stern:2010} & $0.48$ &  $97.0 \pm 62.0$ & \cite{Stern:2010}\\
$0.12$ &  $68.6 \pm 26.2$ & \cite{Zhang:2014} & $0.593$ &  $104.0 \pm 13.0$ & \cite{Moresco:2012} \\
$0.17$ &  $83.0 \pm 8.0$ & \cite{Stern:2010} & $0.68$ &  $92.0 \pm 8.0$ & \cite{Moresco:2012}\\
$0.179$ & $75.0 \pm 4.0$ & \cite{Moresco:2012} & $0.781$ &  $105.0 \pm 12.0$ & \cite{Moresco:2012}\\
$0.199$ &  $75.0 \pm 5.0$ & \cite{Moresco:2012} & $0.875$ &  $125.0 \pm 17.0$ & \cite{Moresco:2012} \\
$0.2$ &  $72.9 \pm 29.6$ & \cite{Zhang:2014} & $0.88$ &  $90.0 \pm 40.0$ & \cite{Stern:2010}\\
$0.27$ &  $77.0 \pm 14.0$ & \cite{Stern:2010} & $0.9$ &  $117.0 \pm 23.0$ & \cite{Stern:2010}\\
$0.28$ &  $88.8 \pm 36.6$ & \cite{Zhang:2014} & $1.037$ &  $154.0 \pm 20.0$ & \cite{Moresco:2012}\\
$0.352$ &  $83.0 \pm 14.0$ & \cite{Moresco:2012} & $1.3$ &  $168.0 \pm 17.0$ & \cite{Stern:2010}\\
$0.3802$ &  $83.0 \pm 13.5$ & \cite{Moresco:2016mzx} & $1.363$ &  $160.0 \pm 33.6$ & \cite{Moresco:2015cya}\\
$0.4$ &  $95.0 \pm 17.0$ & \cite{Stern:2010} & $1.43$ & $177.0 \pm 18.0$ & \cite{Stern:2010}\\
$0.4004$ &  $77.0 \pm 10.2$ & \cite{Moresco:2016mzx} & $1.53$ &  $140.0 \pm 14.0$ & \cite{Stern:2010}\\
$0.4247$ &  $87.1 \pm 11.2$ & \cite{Moresco:2016mzx} & $1.75$ &  $202.0 \pm 40.0$ & \cite{Stern:2010}\\
$0.44497$ &  $92.8 \pm 12.9$ & \cite{Moresco:2016mzx} & $1.965$ &  $186.5 \pm 50.4$ & \cite{Moresco:2015cya}\\
$0.47$ &  $89.0 \pm 49.6$ & \cite{Ratsimbazafy:2017vga}\\
\bottomrule
\end{tabular}
\caption{The CC data used in this work.}
\label{tab:CCdata}
\end{table*}

\section{Systematics}\label{sec.AppendixB}
In this appendix, we review the main systematics involved in the cosmic chronometer and baryon acoustic oscillation data.

\subsection*{Cosmic Chronometers}
In our approach, we have employed SNIa and BAO data to effectively constrain the dimensionless expansion rate $E^2(z)$,  while CC are used to set an absolute scale which provides the value of $H(z)$ at each redshift. The systematics involved in CC measurements could therefore affect our constraints on $H_0$ and $\epsilon$.

Possible sources of systematic errors in these data come from the modelling of the stellar population in the galaxies used as standard clocks~\citep{Moresco:2012by,Moresco:2015cya,Moresco:2016mzx,Moresco:2018xdr,Moresco:2020fbm}.  In particular, a subdominant stellar population may impact the selection of the unbiased tracer, which is required in CC measurements to determine the differential age of the Universe with redshift. Passively evolving galaxies are excellent tracers of the cosmic differential age but the simple description of these galaxies which uses a single stellar population is insufficiently accurate. Instead, the impact of using a realistic star formation history (SFH) must be assessed. The uncertainties in modelling the metallicity of the stellar population must also be considered in the error budget, as this information is used to calibrate the relative age of the population. Finally, the stellar population synthesis (SPS) model used in calibrating the relative stellar age is also a possible source of systematics. 

The CC data employed in the main text (see \autoref{tab:CCdata}) already include the uncertainties coming from the SFH and stellar metallicity~\citep{Moresco:2012by,Moresco:2015cya,Moresco:2016mzx}, while the contribution of subdominant stellar population is negligible for our dataset~\citep{Moresco:2018xdr}. 

The uncertainties associated with SPS models are instead not included. In \cite{Moresco:2020fbm}, these were shown to produce an additional $\lesssim 16\%$ uncertainty in the value of $H(z)$ measured with the CC data. We include this additional error in our fit by first reconstructing the evolution of the SPS contribution with redshift through a GP fit on the values of the third column of Table 3 in ~\citep{Moresco:2020fbm}. We then sum this with that reported in ~\autoref{tab:CCdata}. 

The effect of adding this SPS uncertainty is rather small. We found that it mainly contributes to the $H_0$ uncertainties, leading to $H_0 = 68.3^{+2.9}_{-2.6} $ kms$^{-1}$Mpc$^{-1}$, while the constraint on $\epsilon$ is virtually unchanged, $\epsilon = 0.002^{+0.025}_{-0.029}$. In a recent review by \cite{Moresco:2022phi}, the uncertainties of the CC have been thoroughly checked and the covariance matrix for the SPS modelling error has been derived. Such information about data correlation is however naturally included in the GP reconstruction by assuming a specific kernel shape. In other words, while we only include the diagonal terms of the SPS matrix, their mutual correlation is taken into account through the form of the kernel.  Finally let us note that we tested this statement comparing the full covariance matrix of the SPS errors (following the method proposed in \cite{Moresco:2022phi}) against using only the diagonal terms (as done in this work). We found out that the off-diagonal terms have no significant impact on the $H_0$ uncertainty.
Our conclusions are therefore robust against the inclusion of the SPS modelling error in the CC data.

\subsection*{Baryon Acoustic Oscillations}

For completeness, we lastly compare the results obtained when using the whole BAO dataset to the results obtained by removing the Ly-$\alpha$ data. The high-redshift BAO data (at $ z=2.34 $ and $z= 2.35 $ respectively) peak positions have been shown to have a $\sim 3\sigma$ discrepancy with the CMB prediction from the Planck data in a $\Lambda$CDM model \citep{Aubourg:2014yra,BOSS:2013igd,duMasdesBourboux:2017mrl}. It is therefore important to asses their impact on the results described in the main text.  


An additional consideration is that the GP errors resulting from the fit to the Pantheon data are ill-constrained at higher redshifts, since there are very few SNIa data points above $z \gtrsim 1$. The CC dataset suffers from a similar problem, since its largest redshift is $z=1.965$.

Without including Ly-$\alpha$ data, we found a small shift in the mean value of $H_0$ with almost the same accuracy of the whole BAO dataset, $H_0 = 69.2\pm 2.6 $ kms$^{-1}$Mpc$^{-1}$. The bound on $\epsilon$ instead worsens by a factor of two, $\epsilon=0.004^{+0.040}_{-0.046}$. The discrepancy of the R20 results with $\epsilon =0 $ consequently reduces to $\sim1.5\sigma$ when the R20 prior is combined with our late time data ($\epsilon = 0.031^{+0.041}_{-0.039}$ at 95$\%$ CL). While the R20 discrepancy is slightly reduced, the constraints we draw without the Ly-$\alpha$ BAO do not change the conclusions outlined in the main text. 

\bsp	
\label{lastpage}
\end{document}